\documentclass[12pt,preprint]{aastex}

\slugcomment{Accepted to {\it The Astrophysical Journal}}
\shorttitle{Modeling the System Parameters of 2M\,1533+3759} 
\shortauthors{B.-Q. For et al.}

\newcommand{\gta}{\lower 0.5ex\hbox{$ \buildrel>\over\sim\ $}}
\newcommand{\lta}{\lower 0.5ex\hbox{$ \buildrel<\over\sim\ $}}

\begin{document}

\title{Modeling the System Parameters of 2M\,1533+3759: A New
  Longer-Period Low-Mass Eclipsing sdB+dM Binary}

\author{B.-Q. For\altaffilmark{1}, E.M. Green\altaffilmark{2},
  G. Fontaine\altaffilmark{3}, H. Drechsel\altaffilmark{4},
  J. S. Shaw\altaffilmark{5}, J. A. Dittmann\altaffilmark{2},
  A. G. Fay\altaffilmark{2}, M. Francoeur\altaffilmark{3},
  J. Laird\altaffilmark{2}, E. Moriyama\altaffilmark{2},
  M. Morris\altaffilmark{2}, C. Rodr\'iguez-L\'opez\altaffilmark{6},
  J. M. Sierchio\altaffilmark{2},  S. M. Story\altaffilmark{2},
  A. Strom\altaffilmark{2}, C. Wang\altaffilmark{2},
  S. M. Adams\altaffilmark{2}, D. E. Bolin\altaffilmark{2},
  M. Eskew\altaffilmark{2}, and P. Chayer\altaffilmark{7} } 
\email{biqing@astro.as.utexas.edu}

\altaffiltext{1}{Department of Astronomy, University of Texas, Austin,
  TX 78712}
\altaffiltext{2}{Steward Observatory, University of Arizona, 933 North
  Cherry Avenue, Tucson, AZ 85721}
\altaffiltext{3}{D\'epartement de Physique, Universit\'e
    de Montr\'eal, Montr\'eal, QC H3C 3J7, Canada}
\altaffiltext{4}{Dr. Remeis-Sternwarte Bamberg, Astronomisches Institut 
  der Universit\"{a}t Erlangen-N\"{u}rnberg, Sternwartstra$\beta$e 7,
  96049, Germany} 
\altaffiltext{5}{Department of Physics and Astronomy, University of
  Georgia, Athens, GA 30602}
\altaffiltext{6}{Laboratoire d'Astrophysique de Toulouse-Tarbes,
  Universit\'e de Toulouse, CNRS, 14 Av.\ Edouard Belin, Toulouse 31400,
  France} 
\altaffiltext{7}{Space Telescope Science Institute, 3700 San Martin
  Drive, Baltimore, MD 21218}

\begin{abstract}

We present new photometric and spectroscopic observations for
2M\,1533+3759 (=~NSVS\,07826147), the seventh eclipsing subdwarf B star
+ M dwarf (sdB+dM) binary ever found.  It has an orbital period of
0.16177042 day, or $\sim$3.88~h, significantly longer than the
2.3--3.0 hour periods of the other known eclipsing sdB+dM systems.
Spectroscopic analysis of the hot primary yields $T_{\rm eff} = 29230
\pm 125$~K, $\log g = 5.58 \pm 0.03$ and log $N$(He)/$N$(H) =
$-2.37 \pm 0.05$.  The sdB velocity amplitude is $K_{\rm 1} = 71.1 \pm
1.0$~km~s$^{-1}$.  The only detectable light contribution from the
secondary is due to the surprisingly strong reflection effect, whose
peak-to-peak $BVRI$ amplitudes are 0.10, 0.13, 0.15, and 0.19
magnitudes, respectively.  Light curve modeling produced several
solutions corresponding to different values of the system mass ratio,
$q$ ($M_{\rm 2}$/$M_{\rm 1}$), but only one is consistent with a core
helium burning star, $q$ = 0.301.  The orbital inclination is
$86.6\degr$.  The sdB primary mass is $M_{1} = 0.376 \pm
0.055~M_{\sun}$ and its radius is $R_{1} = 0.166 \pm 0.007~R_{\sun}$.
2M\,1533+3759 joins PG\,0911+456 (and possibly also HS\,2333+3927) in
having an unusually low mass for an sdB star.  SdB stars with masses
significantly lower than the canonical value of $0.48~M_{\sun}$, down
to as low as $0.30~M_{\sun}$, were theoretically predicted by Han et
al.\ (2002, 2003), but observational evidence has only recently begun
to confirm the existence of such stars.  The existence of core helium
burning stars with masses lower than 0.40--0.43~$M_{\sun}$ implies that
at least some sdB progenitors have initial main sequence masses of
1.8--2.0$~M_{\sun}$ or more, {\it i.e.}\ they are at least main sequence 
A stars. The orbital separation in 2M\,1533+3759 is $a = 0.98 \pm
0.04~R_{\sun}$.  The secondary has $M_{2} = 0.113 \pm 0.017~M_{\sun}$,
$R_{2} = 0.152 \pm 0.005~R_{\sun}$ and $T_{\rm eff_{2}} = 3100 \pm
600$~K, consistent with a main sequence M5 star.  If 2M\,1533+3759 
becomes a cataclysmic variable (CV), its orbital period will be 1.6~h, 
below the CV period gap.

\end{abstract}


\section{INTRODUCTION}

Subdwarf B (sdB) stars are evolved, hot, compact stars ($23,000$ K $<
T_{\rm eff} < 37,000$ K; $5.2 < \log g < 6.0$), commonly found in the
disk and halo of our Galaxy \citep{Saffer94}.  They are believed to
ascend the first red giant branch (RGB) following the exhaustion of
central hydrogen, somehow experiencing sufficient mass loss prior to
the RGB tip to remove nearly all of their envelopes. They subsequently
evolve blueward from the RGB before igniting helium in their
cores. From an evolutionary point of view, sdB stars are also known as
extreme horizontal branch (EHB) stars \citep{Heber86}. Their helium
burning cores, generally expected to be just under $0.5~M_{\sun}$, are
essentially identical to those of normal horizontal branch (HB) stars.
However, their hydrogen envelopes are too thin and inert ($<
0.01~M_{\sun}$) \citep{Saffer94,Heber86} to support double shell
burning, so they never make it to the asymptotic giant branch.
Following core helium exhaustion, they evolve directly into sdO stars
before proceeding down the white dwarf cooling track \citep{Dorman93}.

In the context of understanding Galaxy evolution and cosmology, sdB
stars play an important role because their large UV flux appears to be
the dominant source of the ``UV upturn'' phenomenon observed in
elliptical galaxies and the centers of spiral bulges
\citep{DB82,Greg99,Brown97}.  The UV excess in old stellar populations
has been used as an age indicator in evolutionary population synthesis
\citep{Yi97,Yi99}, although more recent work has begun to consider
alternative binary scenarios that would have quite different effects
\citep{P08}.

Various evolutionary scenarios have been proposed for sdB stars, but
the details of the formation mechanisms are not yet well determined.
Possible formation channels can be divided into single star evolution
with enhanced mass loss at the tip of RGB \citep{Castellani93,DCruz96}
and close binary evolution, first suggested by \citet{Mengel76}.
Recently, Han et al.\ (2002, 2003) conducted an in-depth theoretical
investigation through binary population synthesis. They found that
common-envelope evolution, resulting from dynamically unstable mass
transfer near the tip of the first RGB, should produce short-period
binaries ($P \approx 0.1-10$~day) with either a main sequence (MS) or
white dwarf (WD) companion.  If a red giant star loses nearly all of
its envelope prior to the red giant tip via stable mass transfer, a
long-period sdB binary with a MS companion can be produced ($P \approx
10-500$~day).  A most interesting feature of Han et al.'s models is
that they predict a much larger range of sdB progenitor masses than
had previously been considered, including stars sufficiently massive
to avoid a helium flash and instead undergo quiescent helium ignition
in non-degenerate cores (see also \citealp{Hu07};
\citealp{Politano08}).  

Binary formation scenarios appear likely to be responsible for the
majority of observed field sdB stars, as a large fraction are observed
to occur in binaries (e.g., \citealp{Lisker05}; \citealp{MR03};
\citealp{Maxted01}; \citealp{Saffer01}; \citealp{Green97};
\citealp{Allard94}). Nevertheless, the same studies show that there
are a sizable fraction of sdB stars, 30\% or more, that do not now
appear to be in binaries: there is no sign of a companion in high S/N
optical spectra or infrared colors, and their radial velocities are
constant to within the observational errors (a few km s$^{-1}$) over
many months. \citet{MB08} also found a significant fraction, 96\%, of
sdB stars in globular clusters to be single stars, in contrast to
observed field sdB stars. Han et al.\ (2002, 2003) investigated the
possibility of forming single sdB stars by merging two helium white
dwarfs, which would allow the formation of more massive sdB stars
($0.4-0.65~M_{\sun}$), and \citet{Politano08} considered the
possibility that some sdB stars might form from mergers during common
envelope evolution, followed by rotationally-induced mass loss.
Still, unusually high mass loss in single red giant stars cannot yet
be ruled out.

The distribution of sdB masses is clearly one of the most important
constraints on the several possible formation channels.  Different
observational techniques provide different windows of opportunity
for investigating these masses.

More sdB masses have been derived by asteroseismology than by any
other method to date.  Asteroseismology provides an extremely high
level of precision (and is the only way to determine envelope masses,
in addition to total masses), but it has so far been successfully
applied only to the relatively rare short period sdB pulsators.  Two
different types of multimode sdB pulsators have been discovered: short
period V361\,Hya pulsators (originally, EC\,14026 stars,
\citealp{Kilkenny97}) which comprise a rather small percentage of the
hotter sdB stars, and longer period V1093\,Her pulsators (PG\,1716
stars, \citealp{Green03}), which seem to be fairly common among cooler
sdB stars. The rapid oscillations of V361\,Hya stars are interpreted
as low-order pressure modes (p-modes) that are driven by a
$\kappa$-mechanism associated with the radiative levitation of iron in
the thin diffusion-dominated envelopes \citep{Charpinet96,
Charpinet97}.  The same mechanism has also been shown to explain the
excitation of high-order gravity modes (g-modes) in the V1093\,Her
stars \citep{Fontaine03}.  Asteroseismological modeling has so far
been extremely successful with p-mode pulsations in the envelopes of
sdB stars, and the resulting stellar parameters are generally in very
good agreement with theoretical expectations (e.g., 
\citealp{Fontaine08}; \citealp{Charpinet07}, and references therein).
On the other hand, g-mode pulsations, which extend much more deeply
into the stellar cores, will require more sophisticated interior
models before they can be satisfactorily analyzed by asteroseismology
\citep{Randall07}.

The list of p-mode pulsators whose parameters have been derived by
asteroseismology is presented in Table 1.  Most
of the derived masses are within a few hundredths of a solar mass of
the canonical sdB mass of $0.48~M_{\sun}$, except for PG\,0911+456
\citep{Randall07}, which will be discussed further in \S 7.
Interestingly, the only post-common envelope binaries in this list are
Feige\,48 \citep{VanGrootel08a} and PG\,1336$-$018
\citep{Charpinet08}.  Indeed, the large majority of V1093\,Her stars
exhibit low or negligible radial velocity variations, of the order of
a few km s$^{-1}$ or less, and thus must be single stars, or have
extremely low mass companions, or else occur in long period binaries
with a main sequence F, G, or K star primary.  This is not surprising,
since sdB stars whose radial velocity variations are clearly
indicative of post-common envelope binaries are preferentially found
at temperatures cooler than most V1093\,Her stars \citep{Green08}.

Traditional methods of deriving masses by exploiting binary properties
are therefore quite important.  For one thing, binaries provide a
vital test of asteroseismology in the rare cases where the sdB primary
is a pulsator.  More importantly, until improved asteroseismic models
and extensive satellite observations make it possible to successfully
model g-mode sdB pulsators, the only way to derive masses
for a larger sample of post-common envelope sdB stars is to analyze
their binary properties.  Finally, there are simply a large number of
binaries that contain non-pulsating sdB stars.

The difficulty with most sdB stars in post-common envelope systems is
that they are single-lined spectroscopic binaries with essentially
invisible compact secondaries.  In principle, precise measurements of
the sdB surface gravity and rotational velocity in a tidally locked
system will yield the orbital inclination, allowing the individual
component masses to be determined from the mass function (e.g.,
\citealp{Geier08}), but the accuracy of this approach has not yet been
proven. There are, however, a small number of rare post-common
envelope sdB+dM binaries \citep{Maxted04}, which have been identified
by their reflection effects -- e.g., the sinusoidal variation
observed in the light curve due to reradiated light from the heated
side of the tidally locked M dwarf -- that are more promising.  The
known sdB+dM systems are summarized in Table 2.  If
narrow lines originating from the cool secondary could be detected,
then masses of both components could be derived from the double-lined
spectroscopic solution.  Again, this should be possible in principle,
especially in binaries with the shortest orbital periods, where the
heated face of the secondary is brighter than it otherwise would be,
but results so far have been ambiguous.  \citet{Vuckovic08} detected
emission lines from the secondary in PG\,1336-018, by subtracting the
spectrum of the hot primary from spectra taken at other phases, but
the S/N of the spectra were only sufficient to claim general
consistency with the orbital solution described in \citet{Vuckovic07}.
Using much higher S/N spectra of a similar sdO+dM binary, AA~Dor,
\citet{Vuckovic08} were able to determine a velocity amplitude for the
secondary, but their derived primary mass has now been vigorously
disputed by \citet{Rucinski09}.  \citet{WS99} presented a good
argument for the detection of $H\alpha$ absorption lines from the
secondary in HW~Vir, again by subtracting the spectrum near minimum
light from spectra near maximum light, and obtained
reasonable velocities, but it is perplexing that absorption lines and
no emission lines should have been seen.

An apparently more successful method is to model the light variations
in sdB+dM binaries exhibiting reflection effects, especially the
eclipsing systems, in order to determine the system parameters.  This
is a very complex endeavor.  The models have many free parameters, and
there are large uncertainties that typically require additional
information to constrain the solution.  Often, the light curves
provide more than one high quality solution.  For example,
\citet{Drechsel01} had to make use of a mass--radius relation for the
secondary star to decide between two solutions that implied quite
different sdB masses for HS\,0705+6700 (0.483 and $< 0.3~M_{\sun}$).
\citet{Heber04} needed to use their spectroscopic $\log g$ and
mass--radius relations to discriminate between two solutions with
different secondary albedos and inclinations in HS\,2333+3927.
\citet{Vuckovic07} found three possible solutions modeling the light
curves PG\,1336$-$018, and it was not possible to choose between two
of them until \citet{Charpinet08} derived a consistent primary mass by
asteroseismological modeling.  Furthermore, even when a single family
of solutions can be identified, there still remain unavoidable
ambiguities in choosing one ``best'' model \citep{Drechsel01}.  Even
in the most favorable cases of eclipsing sdB+dM binaries, the eclipses
are not flat-bottomed, leading to a small range of nearly equivalent
solutions in the vicinity of the deepest minimum.  The resulting small
variations in the mass mass ratio, $q$, lead to a significant range in
the derived sdB mass.  The uncertainties are obviously larger when
there is no eclipse.  Still, light curve modeling provides valuable
information, and when the derived sdB mass can be verified -- rarely
by asteroseismology, more often from consistency with the
spectroscopic surface gravity or projected rotational velocity --
our confidence in the results is greatly increased.  It is clearly
important to investigate as many sdB+dM binaries as possible,
especially the eclipsing systems, in order to build up a more
comprehensive picture of sdB masses produced by post-common envelope
evolution and to compare with the distribution of masses from other
formation channels.

In this paper, we report on the system parameters of 2M\,1533+3759
($15^{h}33^{m}49.44^{s}$, $+37\degr59\arcmin28.2\arcsec$, J2000), a
new eclipsing sdB+dM binary with a longer orbital period than any
eclipsing sdB+dM discovered so far.  This star was first recognized as
an sdB in 2005 (although it remained unpublished) during a
continuing spectroscopic survey \citep{Green08} of bright blue stellar
candidates selected from a variety of sources, including the 2MASS
survey \citep{Skrutskie06}.  The current investigation was motivated
by \citet{KS07}, who discovered that 2M\,1533+3759 is an eclipsing
binary, NSVS\,07826147, through their work with the Northern Sky
Variability Survey (NSVS; \citealp{Wozniak04}).  \citet{KS07}
identified a group of nine eclipsing binaries with short periods and
relatively narrow eclipse widths, indicating very small radii for the
components.  Since their list includes the well-known HW~Vir
(\citealp{Lee09} and references therein), as well as 2M\,1533+3759,
which we confirmed to be a spectroscopic near-twin of HW~Vir,
\citet{KS07} proposed that the other objects in their Table 3
might also be sdB+dM binaries.  \S 2 presents the results from our
follow-up spectra for these stars.

In \S 3, we describe new spectroscopy and photometry for
2M\,1533+3759.  The data analyses are given in \S 4 and \S 5, and the
system parameters are derived in \S 6.  We discuss possible selection
effects and consider the unusually low derived mass for the sdB mass
in \S 7.  \S8 looks at the evolution of 2M\,1533+3759, and \S 9
contains our conclusions.

\section{NSVS ECLIPSING SDB+DM CANDIDATES}

We have obtained high S/N low resolution spectra for Kelley \& Shaw's
(2007) proposed sdB+dM stars (their Table~3).  All were observed with
the same telescope and instrumental setup (\S 3) that we used to
obtain our initial spectrum of 2M\,1533+3759.

Table 3 of this paper presents the results of our
spectroscopic follow-up.  The NSVS numbers, $V$ magnitudes and orbital
periods from \citeauthor*{KS07} are listed in the first three columns.
Columns 4, 5, and 6 give the $J-H$ color, RA, and Dec from the 2MASS
All-Sky Point Source Catalog \citep{Skrutskie06} for the objects that
we observed.  The seventh column lists our best estimate of their
spectral types.  For the non-sdB stars, the spectral types were
  determined by cross-correlating their continuum-subtracted spectra
with template spectra of known main sequence spectral standards
\citep{GC09}, acquired with the same instrument and spectroscopic
setup, in order to find the best match.  Since the binary spectra
are composite, the best matches indicate either the dominant or the
effective spectral type.

NSVS\,04818255 deserves further comment. Its NSVS coordinates are
$08^{h}40^{m}59.8^{s}$, $+39\degr56\arcmin02\arcsec$; this is close,
but not quite coincident with the brightest star in the immediate
area.  \citeauthor*{KS07} identified NSVS\,04818255 with the sdB star
PG\,0837+401.  However, according to the finder chart in
\citet{GSL86}, PG\,0837+401 is the fainter star at
$08^{h}41^{m}01.3^{s}$, $+39\degr56\arcmin18\arcsec$, approximately
$24\arcsec$ northeast; our spectrum confirms that it is indeed an sdB
star.  We initially observed the bright F9--G0 star nearest to the
NSVS coordinates, since it has the same 2MASS $J-H$ value that
\citeauthor*{KS07} give for NSVS\,04818255.  However, S.\ Bloemen and
I.\ Decoster (Leuven) and M.\ Godart (Li\`ege) recently obtained
time-series photometry indicating that neither PG\,0837+401 nor the
bright F9--G0 star are variable \citetext{R.\ {\O}stensen, priv.\
comm.}.  The eclipsing system that they identify with NSVS\,04818255
is the intermediate brightness object almost $40\arcsec$ west
northwest of PG\,0837+401.  We obtained a spectrum for the variable
star and found it to have a G0 spectral type, in agreement with its
somewhat redder $J-H$.

HW~Vir and 2M\,1533+3759 are therefore, unfortunately, the only two
bonafide sdB stars in Kelley \& Shaw's (2007) list.  Figure 1
shows our flux-calibrated spectrum for 2M\,1533+3759, along with the
bluest and reddest of the non-sdB spectra from Table 3, for
comparison.  It is clear from the decreasing flux blueward of the
Balmer jump that there aren't any sdB stars hidden in any of the seven
binaries with overall A, F, or G spectral types.  $J-H$ colors are a
good indicator for the presence of an sdB star in a suspected sdB+dM
binary, since M dwarfs later than about M2 are too faint relative to
sdB stars to have much of an effect on the $J-H$ colors.  All of the
known sdB+dM binaries have $-$0.2 $<$ $J-H$ $<$ 0.0; their distribution
in $J-H$ is only slightly redder than the overall distribution of
moderately unreddened sdB+WD binaries and non-binary sdB stars plotted
in Green et al.'s (2008) Figure~5.

\section{OBSERVATIONS AND REDUCTIONS}

\subsection{Spectroscopy}

Low-resolution spectra for 2M\,1533+3759 were obtained with the Boller
\& Chivens (B\&C) Cassegrain spectrograph at Steward Observatory's
2.3~m Bok telescope on Kitt Peak. The 400 mm$^{-1}$ first order
grating was used with a 2.5\arcsec\ slit to obtain spectra with a
typical resolution of 9~\AA\ over the wavelength interval
3620--6900~\AA.  The instrument rotator was set prior to each
exposure, to align the slit within $\sim 2\degr$ of the
parallactic angle at the midpoint of the exposure.  HeAr comparison
spectra were taken immediately following each stellar exposure.  The
spectra were bias-subtracted, flat-fielded, background-subtracted,
optimally extracted, wavelength-calibrated and flux-calibrated using
standard IRAF tasks.  Details of the individual low resolution spectra
are given in Table 4. The orbital phases in the last column
are discussed in \S 5.1.

We acquired additional medium resolution spectra in 2008 and 2009 for
radial velocities, again with the B\&C spectrograph on the 2.3~m Bok
telescope.  For these, we used an 832 mm$^{-1}$ grating in second
order with a 1.5\arcsec\ slit to achieve 1.8~\AA\ resolution over a
wavelength range of 3675--4520~\AA.  The slit was aligned with the
parallactic angle at the midpoint of each exposure, the same as for
the low resolution spectra, but comparison HeAr spectra were taken
before and after each stellar spectrum. The spectra were reduced in
a similar manner, except that they were not flux-calibrated.  After
wavelength calibration, the radial velocity spectra were interpolated
onto a log-wavelength scale.  The continuum was removed from each
spectrum by dividing through by a spline fit to the continuum, and
then subtracting a constant equal to unity in order to get a continuum
value of zero.  Table 5 lists the details of the medium resolution
spectra.  The radial velocities are described in \S 4.1 and the
orbital phases in \S 5.1.

\subsection{Differential Photometry}

Differential $BVRI$ light curves for 2M\,1533+3759 were obtained at
the Steward Observatory 1.55~m Kuiper telescope on Mt.\ Bigelow,
Arizona, between February and June of 2008 and in March 2009.  We used
the Mont4K facility CCD camera\footnote{See
http://james.as.arizona.edu/~psmith/61inch/instruments.html for a
description of the Mont4K CCD imager and filters.} with Harris $BVR$
and Arizona $I$ filters. Several hundred bias images and dome flats
were obtained each day to reduce the error budget due to calibrations
to less than 0.001 magnitude. The time stamp for each image is written
by the clock on the CCD computer, which is synchronized with the on-site
GPS system every 120~s, so that the times are always correct to better
than a couple of tenths of a second.  To reduce the observational
sampling time, we used on-chip $3\times3$ binning and read out only
2/3 of the CCD rows, resulting in a readout time of 22~s per image.
(For 2009, the readout time was reduced to 14~s, as a result of
improvements to the electronics.)  The remaining overhead time between
images was 7~s, including 6~s for the filter change.  We alternated
between two filters each night in order to obtain two coeval light
curves while maintaining adequate sampling of the eclipses.
Table 6 summarizes the photometric observations.

The images were reduced with a pipeline constructed from standard IRAF
tasks. The bias-subtracted images were flat-fielded with the
appropriate $BVRI$ dome flat and corrected for bad columns and cosmic
rays.  Images in the $I$ filter were further corrected by subtracting
a scaled, high S/N, zero-mean fringe frame.  The fringe frame was
constructed from 31 dithered $I$ images, 600~s each, in fields with
low stellar density, taken between March and May 2008; the fringe
pattern was very stable over that time interval.  Aperture photometry
was performed for the sdB and a set of reference stars, with the
aperture radius set to 2.25 times the average FWHM in each image.  The
same set of eight, apparently nonvariable, reference stars was used
with every filter; the reference stars were chosen to be distributed
as closely and symmetrically as possible around 2M\,1533+3759
(Fig. 2).  The differential magnitudes (sdB minus the average of the
reference stars) were converted to relative fluxes and normalized to
1.0 near the quarter phase of the star's orbit.

The resulting light curves, shown below in Figure 6 and
further discussed in \S7, have well-defined primary and secondary
eclipses.  The peak-to-peak amplitudes of the reflection effect are
0.10, 0.13, 0.15, and 0.19 magnitudes, respectively, in the $BVRI$
filters.

\section{SPECTROSCOPIC ANALYSIS}

\subsection{Radial Velocities}

We derived the radial velocities iteratively using a double-precision
version of the IRAF task FXCOR.  The initial velocity template was
fadsconstructed by combining and median-filtering all 38 medium resolution
spectra.  The individual spectra were cross-correlated against the
template by fitting a gaussian to the cross-correlation peak to
determine the velocity shifts.  The spectra were then Doppler-shifted
to the same velocity and recombined into an improved template.
Five iterations were required to reach convergence.  Columns 5 and 6
in Table 5 list the derived radial velocities and their
associated errors.  Since FXCOR velocity errors are only known to
within a scale factor, the final step was to scale the FXCOR errors so
that the average error matches the standard deviation of the observed
points about the fitted velocity curve.

The radial velocity solution was determined using a weighted
least-squares procedure to fit a sine curve. The orbital period was
fixed at the value derived from the eclipse times in the following
section, since the photometric period is much more precise than the
period derived from the velocities.  The radial velocity solution is
shown in Figure 3.  The velocity semi-amplitude is $K_{\rm 1}
= 71.1 \pm 1.0$ km s$^{-1}$.  The systemic velocity, $\gamma = -3.4
\pm 5.2$ km s$^{-1}$, was determined relative to three sdB radial
velocity ``standards'', PG\,0101+039, PG\,0941+280, and PG\,2345+318,
one or two of which were observed each night\footnote{These are
actually short-period sdB+WD binaries with large velocity amplitudes
that we have observed for 10 to 15 years, whose velocities
are known to 1--2~km s$^{-1}$ at any given time.}.

\subsection{Spectroscopic Parameters}

We fit the Balmer lines from H$\beta$ to H11 and the strongest helium
lines (4922~\AA, 4471~\AA, and 4026~\AA) in our low resolution spectra
to synthetic line profiles calculated from a grid of zero metallicity
NLTE atmospheric models.  Our expectation was that the reflection
effect in 2M\,1533+3927 would introduce negligible contamination from
the secondary.  The only sdB+dM binary whose spectroscopic parameters
have previously been reported to vary with orbital phase is
HS\,2333+3927 \citep{Heber04}, and its reflection effect is more than
twice as large as that of 2M\,1533+3927.  We were therefore surprised
to find that our individual low resolution spectra for 2M\,1533+3927
do in fact give significantly different temperatures at different
orbital phases, amounting to the better part of 1000~K.

We therefore returned to our more numerous medium resolution spectra,
and (after reinterpolating onto a linear wavelength scale) fit
H$\gamma$ through H11, He 4471~\AA, and 4026~\AA, again using zero
metallicity NLTE models.  The medium resolution spectra show the same
orbital temperature effect (Fig. 4), with about the same
amplitude, even though they exclude H$\beta$ (which suffers the most
from contamination by the secondary of all the lines we considered).
The lowest derived temperatures are found from spectra taken near
minimum light.  The unexpected prominence of the temperature
variations with orbital phase is probably due to the high S/N noise of
our spectra (70--90 per pixel).  There is also a suggestion of a
similar trend with gravity, but the derived helium abundances were
negligibly affected.  (For unknown reasons, our temperature variations
are in the same sense as those derived by \citet{Heber04} using only
helium lines (their Fig.~7b), and in the opposite sense from what they
found when fitting both Balmer and helium lines, although naturally we
see smaller amplitude variations for 2M\,1533+3759.)

To be safe, we adopted atmospheric parameters determined from 14
spectra observed near minimum light, i.e., orbital phases
between 0.8 and 1.2, not including the two points closest to the
center of the eclipse.  (The temperature derived at the midpoint of
the primary eclipse was surprisingly discrepant, possibly due to
absorption of some of the uneclipsed sdB light near the limb of the
secondary; discrepant gravity values were also seen during both
eclipses.)  The excellent quality of the fit can be seen in
Figure 5.  Our adopted spectroscopic parameters are
$T_{\rm eff} = 29230 \pm 125$~K, $\log g = 5.58 \pm 0.03$, and
log $N$(He)/$N$(H) = $-2.37 \pm 0.05$, where the errors are the
standard deviations of the values from the individual spectra.  This
$T_{\rm eff}$ was used as the initial value for the primary
temperature in our light curve modeling in \S 5.2.

\section{PHOTOMETRIC ANALYSIS}

\subsection{Ephemeris}

We solved for the orbital period using a linear least-squares fit to the
well-defined times of primary and secondary eclipse minima in the $V$
and $R$ light curves, in the equation $T_{min} = T_{0} + nP$,
 where
 $T_{min}$ are the times of the eclipse minima,
 $T_{0}$ is the reference HJD for the primary eclipse at $n$ = 0,
 $n$ are the cycle numbers, and
 $P$ is the orbital period in units of a day.

The time of minimum for each observed primary and secondary eclipse
was determined by fitting an inverse Gaussian to the eclipse shape.
The results are listed in Table 7, along with the
corresponding cycle numbers, the instrumental filter, and the $O-C$
time residuals.  The standard deviation of the $O-C$ values is 3.3~s.
The derived ephemeris for the primary eclipses is

\centerline{HJD = (2454524.019552 $\pm$ 0.000009)$+$(0.16177042 $\pm$
  0.00000001)$\times$ E.} 

\subsection{Light Curve Modeling}

The $BVRI$ light curves were phased with the ephemeris and orbital
period derived from the photometry.  Small vertical flux differences
equivalent to a few hundredths of a magnitude remained in the phased
light curves.  These could be due to slight long term variability in
one or more of the reference stars, but are more likely to be caused
by subtle variations in the dome flats from different runs.  We
therefore shifted the light curves in the same filter vertically by a
small constant to minimize the standard deviation of the total phased
light curves for that filter.  The light curves for all four filters
were analyzed simultaneously with the MORO (MOdified ROche) code
\citep{Drechsel95}.

The MORO code adopts the Wilson-Devinney monochromatic light,
synthetic light curve calculation approach (\citealp{WD71}; hereafter
WD), but has implemented a modified Roche model that takes into
account radiation pressure effects in close binaries with hot
components.  It also replaces the classical WD grid search
differential corrections method with a more powerful $SIMPLEX$
optimization algorithm.  This provides several advantages: in
particular, the fitting procedure improves with each iteration and is
not allowed to diverge.  For details of the numerical procedure and
the radiation pressure implementation, we refer the reader to the
description in \citet{Drechsel95}.

Light curve modeling becomes a challenging task when information about
the secondary is limited, as is the case in all single-lined
spectroscopic binaries.  Since the modeling requires a large set of
parameters, it is important to constrain as many as possible based on
additional spectroscopic and theoretical information.  We assumed the
orbit is circular and the stellar rotation is synchronized with the
orbit, since the time scales for both circularization and
synchronization are a few decades \citep{Zahn77}, very much shorter
than the helium burning lifetime of a horizontal branch star.  We
adopted the spectroscopic $T_{\rm eff}$ of the sdB as an initial
parameter, and took the linear limb-darkening coefficients ($x_{1}$)
of 0.305, 0.274, 0.229 and 0.195 from \citet{DC95} and \citet{WR85}
for the $B$, $V$, $R$, and $I$ filters, respectively. These values
correspond to the nearest available stellar atmosphere model, a star
with $T_{\rm eff} = 30,000$~K and $\log g = 5.0$, and should be very
close to the correct values \citep{Wood93}, since the dependence on
the surface gravity is weak.  Previous experience with light curve
modeling of similar systems \citep{Hilditch96} indicates that the
limb-darkening coefficient of the cool secondary star ($x_{2}$) can
deviate highly from normal values for cool dwarf stars, so we decided
to treat $x_{2}$ as an adjustable parameter. Due to the irradiation
effect, the limb-darkening can be expected to be more extreme than for
single stars, and thus we employed initial values of 0.7, 0.8, 1.0 and
1.0, for the $B$, $V$, $R$, and $I$ filters, respectively.  The
primary albedo ($A_{\rm 1}$) was fixed to 1.0 and its gravity
darkening exponent ($\beta_{1}$) was set to 1.0, appropriate for a
radiative outer envelope \citep{vonzeipel24}.  The enormous
reflection effect suggests a mirror-like surface on the heated side
facing the primary, indicating complete reradiation of the primary
light; therefore a secondary albedo ($A_{2}$) of 1.0 was adopted.  We
set the gravity darkening exponent ($\beta_{2}$) to 0.32 for the
convective secondary \citep{Lucy67}.  The radiation pressure parameter
for the secondary star ($\delta_{2}$) was set to zero because the
radiation pressure forces exerted by the cool companion are
negligible.  A blackbody approximation was used to treat the
irradiation of the secondary by the primary.  We input central
wavelengths of 4400, 5500, 6400, and 7900\AA\ for our $BVRI$
passbands, which are a fair match to the filter passbands convolved
with the CCD sensitivity.

The simultaneous light curve modeling was performed with the WD mode 2
option, for a detached system.  The remaining free parameters for the
fitting procedure include the orbital inclination, $i$; the effective
temperature of the secondary, $T_{2}$; the Roche surface potential,
$\Omega_{1}$ and $\Omega_{2}$; the mass ratio, $q = M_{2}/M_{1}$; the
color-dependent luminosity of the primary, $L_{1}$; the radiation
pressure parameter for the primary, $\delta_{1}$; and $l_{3}$, a
potential third light contribution due to a possible unresolved field
star or an extended source. The color-dependent luminosity of the
secondary, $L_{2}$, was not adjusted but was recomputed from the
secondary's radius and temperature.

Degeneracy is a common problem encountered in light curve modeling.  A
high degree of correlation between several parameters (e.g., 
$i$, $q$) can result in several equally good solutions with different
families of parameters.  Therefore, it is necessary to test for the
presence of multiple good solutions over a wide range of mass ratios.
The usual procedure is to run a series of initial trials at discrete
mass ratios, keeping them fixed.  Unfortunately, our first set of
trials did not produce any good solutions for mass ratios in the range
$1.2 < q < 0.2$, corresponding to either an sdB mass of
0.49$~M_{\sun}$ and M dwarf masses in the range 0.6 to 0.1$~M_{\sun}$
(M0 to M5.5), or to smaller sdB masses and later M spectral types --
i.e., there were no solutions that matched the shapes of our
observed light curves -- because the reflection effect was
underestimated by about 30$\%$ in all of the models.  The trial runs
did however suggest that there was no third light contribution, so we
set that parameter to zero for the rest of the runs.

A similar, although less extreme, problem was encountered in previous
attempts to model the light curves of eclipsing sdB+dM binaries
(\citealp{Kilkenny98}, PG\,1336$-$018; \citealp{Drechsel01},
HS\,0705+6700), especially with redder filters, and for the same
reason: theoretical models aren't sophisticated enough in their
treatment of the reflected/reradiated light. Both \citet{Kilkenny98}
and \citet{Drechsel01} found that if the secondary albedo was treated
as a free parameter, their solutions converged to physically
unrealistic values, $A_{2} > 1.0$, although they were able to find
acceptable solutions when $A_{2}$ was held fixed at a value of
1.0. \citet{Vuckovic07} and \citet{Lee09}, both using Wilson-Devinney
synthesis codes, noted that their biggest difficulty concerned the
temperature of the heated secondary.  This appears to be an alternate
version of the same basic problem, i.e. correctly treating the
light from the secondary star, which manifests differently in
different adaptations of the WD code.  \citet{Vuckovic07} was able to
find good solutions with $A_{2} = 0.92$ by simply fixing their
secondary temperature at the average of the values found separately in
their two passbands. \citet{Lee09} had to resort to mode 0 instead of
mode 2, allowing $L_{2}$ and $T_{2}$ to be separate free parameters
(rather than computing $L_{2}$ from $T_{2}$ and $R_{2}$), in addition
to fixing $A_{2} = 1.0$.  Since we could not find any acceptable fits
to our light curves with MORO when $A_{2}$ was set to 1.0, we
decided to treat it as an adjustable scale factor, accepting that
it would converge to an unphysically high value.

When $A_{2}$ was no longer kept fixed, good fits to the light curve
shapes were found for the following mass ratios: $q = $0.301, 0.586,
0.697, 0.800, and 0.888.  To discriminate between the possible
solutions, we calculated the sdB mass corresponding to each value of
$q$, using the mass function, which can be expressed as

\begin{equation}
\frac{M_{1}\times(q~{\rm sin}~i)^{3}}{(1+q)^{2}} = \frac{K_{1}\,^{3}
  P}{9651904},
\end{equation}

\noindent where $i$ is the corresponding inclination angle, which was
always $86.6\degr \pm 0.2\degr$, and with $K_{\rm 1}$ = 71.1 km s$^{-1}$
and $P$ = 0.16177042 day, as derived above.  The resulting sdB masses
are 0.376, 0.076, 0.052, 0.038, and $0.031~M_{\sun}$, respectively.
According to evolutionary models, core helium burning sdB stars must
have masses substantially larger than $0.08~M_{\sun}$, leaving only
one reasonable solution, $q$ = 0.301.

Once $q$ was constrained to a single approximate value, the problem
was reduced to finding the deepest minimum in the surrounding
multidimensional parameter space.  The $SIMPLEX$ algorithm is a very
powerful numerical tool, but it is always possible for any algorithm
to converge into a less-than-optimal local minimum.  To verify that
the converged $q = 0.301$ solution was the deepest minimum in the
local vicinity, we varied the set of starting parameters over $0.27 <
q < 0.35$ ($0.26 < M_{1} < 0.50~M_{\sun}$) in multiple additional
runs, to make sure that they all converged to the same solution within
a small error margin, which they did.  Table 8 lists the
best light curve solution for 2M\,1533+3759 for all the filters.  The
standard deviations of the various fits are at the bottom.  The
observed $BVR I$ light curves are shown together with the calculated
theoretical curves in Figure 6.

Throughout the previous runs, the temperature of the primary, $T_{1}$,
was initialized to the spectroscopic value, but it was allowed to be
an adjustable parameter.  The converged results showed a consistent
preference for a higher-than-observed effective temperature, by 1200~K
or so.  However, once we isolated the best model, we reran the
solution while keeping $T_{1}$ fixed at 29230\,K. The resulting
values of the mass ratio, inclination angle, fractional radii, etc.,
in Table 8 are the same, within the errors, whether
$T_{1}$ is 30400\,K or 29230\,K.

Figure 7 is a series of snapshots from a 3D-animation of 2M\,1533+3759
at different orbital different phases. 

\section{GEOMETRY AND SYSTEM PARAMETERS}

The light curve solution allows us to calculate the absolute system
parameters.  Substituting the values of $K_{\rm 1}$ and $P$ from \S
4.1 and \S 5.1 into Eq.~(1), along with $q = 0.301$ and $i =
86.6\degr$, results in component masses $M_{\rm 1}$ = $0.376 \pm
0.055 ~M_{\sun}$ and $M_{\rm 2}$ = $0.113 \pm 0.017 ~M_{\sun}$.
Kepler's law tells us the orbital separation of the two stars, $a =
0.98 \pm 0.04~R_{\sun}$, which can then be used to scale the
fractional radii from the model
solution in order to get the actual radii, $R_{\rm 1}$ = $0.166 \pm
0.007~R_{\sun}$ and $R_{\rm 2}$ = $0.152 \pm 0.005 ~R_{\sun}$.

The light curve modeling is completely independent of the observed
spectroscopic gravity, which therefore provides a nice consistency
check.  The calculated $\log g$ corresponding to our derived $M_{\rm
1}$ and $R_{\rm 1}$ turns out to be $5.57 \pm 0.07$, essentially
identical with our adopted spectroscopic value of 5.58.

In the past, error bars have not usually been attached to masses
derived from modeling light curves of sdB+dM binaries, but we found it
to be a very instructive exercise.  The formal error propagation for
the primary mass, according to equation (1), includes the uncertainties on
$q$, $i$, and $K_{\rm 1}$, and $P$.  Although the mass depends on the
cubic power of both $K_{\rm 1}$ and $q$, the error in $K_{\rm 1}$ is
small enough in our case that the mass uncertainties are dominated by
the uncertainty in $q$, as small as it is. 95\% of the error in $M_{\rm
1}$ is due to the $3 M_{\rm 1} \Delta q / q$ term.  Our inability to
more tightly constrain the sdB mass is a dramatic illustration of why
useful mass constraints from light curve modeling can usually be
obtained only for eclipsing systems (unless, of course, good radial
velocities can be obtained from both components).  Furthermore, even
with an eclipsing sdB+dM binary, the light curve shapes and velocity
amplitude must be sufficiently precisely observed to adequately
minimize the other error terms, or else the uncertainty in the mass
will be even larger.

The temperature of the secondary is somewhat more uncertain, $3100 \pm
600$~K, since it contributes almost negligibly to the total light,
aside from the reflection effect.  Nevertheless, our model value for
$T_{2}$ is quite acceptable.  According to the theoretical
$T_{\rm eff}$ -- mass -- luminosity relation of \citet{BC96}, the
predicted temperature and radius of a $0.113~M_{\sun}$ main sequence
star should be 2854\,K and $0.138~R_{\sun}$, respectively,
corresponding to an M5 dwarf.  The empirical mass -- radius relation of
\citet{B006} for low mass main sequence stars gives an identical 
radius of 0.138~$R_{\sun}$.  Our value of 0.152~$R_{\sun}$ is slightly
larger (although still within the 3$\sigma$ error), but it would not
be unexpected if the highly heated and already slightly distorted
secondary in a system like 2M\,1533+3759 turned out to be a little
larger than an isolated M dwarf of the same mass.

Table 9 summarizes the system parameters for
2M\,1533+3759, beginning with our adopted spectroscopic
parameters and the photometric and radial velocity solutions
described in the previous sections.

\section{DISCUSSION}

We examined several possible systematic effects, begining with our
spectroscopic parameters.  Under the reasonable assumption that the
primary's rotation is synchronized with the orbital period, its
rotational velocity should be $V_{\rm rot_1} = 2\pi R_{1} / P = 52 \pm
2$ km~s$^{-1}$.  This corresponds to 1.0 pixel in our medium
resolution spectra, which have an instrumental FWHM of 2.75 pixels.
We reanalyzed our combined minimum-light spectrum after broadening the
synthetic spectra by this extra amount, and found that the expected
rotation has a negligible effect on the spectroscopic parameter
determination.  The derived temperature was reduced by 10\,K and the
gravity was reduced by 0.002 dex.

Next, we investigated the effects of using zero metallicity NLTE
atmospheres to derive our spectroscopic parameters, since metal lines
are observed to be present in sdB atmospheres, especially in the UV.
Two of us (G.F. and P.C.) conducted an experiment in which TLUSTY was
used to construct a synthetic model atmosphere at a temperature of
28000\,K, $\log g = 5.35$, log $N$(He)/$N$(H) = $-$2.70, and solar
abundances of C, N, O, S, and Fe.  Using our zero metallicity NLTE
grid, the derived parameters were found to be $T_{\rm eff} =
30096$\,K, $\log g = 5.54$, and log $N$(He)/$N$(H) = $-$2.72.  At
these abundances, we would have overestimated the effective
temperature by about 2000\,K and the surface gravity by almost 0.2
dex, so the true values for 2M\,1533+3759 would be about 27300\,K and
5.40, respectively.  Happily, the light curve solution is amazingly
robust.  The model results obtained by further lowering the primary
temperature to a fixed value of 27300\,K are only negligibly different
from our original solution.  Thus, the system parameters would
remain essentially the same: $q = 0.303$, $i = 86.5\degr$, $M_{\rm 1}
= 0.370~M_{\sun}$, $M_{\rm 2} = 0.112~M_{\sun}$, $R_{\rm 1} =
0.165~R_{\sun}$, $R_{\rm 2} = 0.152~R_{\sun}$, and $a =
0.98~R_{\sun}$.  The calculated sdB surface gravity would also be
unchanged, $\log g = 5.57 \pm 0.03$, but would no longer be as
consistent with the expected gravity of 5.40. This implies that the
atmospheric abundances in 2M\,1533+3759 are not as large as the solar
values assumed above.

We spent considerable time worrying about the very large secondary
albedo, $A_{2} \sim 2$, that was required to obtain a solution which
fits the observed shapes of the 2M\,1533+3759 light curves, since all
previous sdB+dM analyses were able to find acceptable light curve
solutions with $A_{2} \sim 1$.  We tested the version of MORO running
at the University of Texas using Drechsel et al.'s (2001) input
datafile, and found exactly the same solution that they did.  We
verified that an independent Steward $V$ light curve data for
HS\,0705+6700, in the same format as our 2M\,1533+3759 data, produced
a curve that fell exactly between Drechsel et al.'s (2001) normalized
$B$ and $R$ data for HS\,0705+6700, thus eliminating problems with our
input format.  We shifted the $BVRI$ effective wavelengths specified
to MORO by up to 200~\AA, with no effect on the output solution.

Our dataset is unique among published sdB+dM light curve analyses in
extending to the $I$ filter.  \citet{Drechsel01} fit only $B$ and $R$
data, \citet{Heber04} fit $BVR$, \citet{Vuckovic07} used $g^{\prime}$
(intermediate between $B$ and $V$) and $r^{\prime}$ (close to $R$),
and \citet{Lee09} had only $V$ and $R$.  We therefore reanalyzed our
2M\,1533+3759 data using only the $B$ and $R$ light curves. The
results were the same as before: when $A_{2}$ is allowed to be a free
parameter, the solution always converges to $A_{2}$ near 2. 
Furthermore, no new solutions appear for other $q$ values, and the
solution for $q = 0.301$ is nearly identical to our previous best
solution.  If $A_{2}$ is forced to have a value of 1, the $B$ and $R$
solutions fail to fit the observed light curve shapes in nearly the
same manner as our original trial solutions at the same $A_{2}$ and
$q$.  The amplitude of the theoretical reflection effect with $A_{2} =
1$ using current models simply isn't large enough to fit
2M\,1533+3759.

An alternate way to look at this problem is to compare the reflection
effect amplitudes in 2M\,1533+3759 versus\ HW~Vir.  HW~Vir was selected
because it has the next longest orbital period of well-studied eclipsing
sdB+dM systems besides 2M\,1533+3759, and because our high S/N spectra
give essentially identical temperatures and gravities for these stars
when analysed in a homogeneous manner.  However complicated the
physics of the reflection effect may be, the actual processes ought to
be similar in both systems.  Thus, to first order, the reflection
effect amplitudes should be proportional to the luminosity of the
primary and the surface area of the heated face of the secondary, and
inversely proportional to the distance between the two stars.  Using
our values of $R_{\rm 1}$, $T_{{\rm eff}\,1}$, $R_{\rm 2}$, and $a$ for
2M\,1533+3759, and Lee et al.'s (2009) values for HW~Vir
($0.183~R_{\sun}$, 28490\,K, $0.175~R_{\sun}$, and $0.86~R_{\sun}$,
respectively) to calculate the ratio of $R_{\rm 1}^2 T_{{\rm eff}\,1}^4
R_{\rm 2}^2 / a^2$ for the two binaries, we find that the amplitude in
2M\,1533+3759 ought to be 53\% of the amplitude in HW~Vir. Instead, it
is observed to be 95\% of the HW~Vir amplitude.  It seems that the
reflection effect in 2M\,1533+3759 really is stronger than would be
expected, compared to other known eclipsing sdB+dM binaries.  Another
light curve solution might give a different result, but an exhaustive
search of parameter space failed to find any other solution that fit
our data.

The most interesting result of our modeling is the unusually low mass
obtained for the sdB star in 2M\,1533+3759.  The vast majority of sdB
masses derived previously from asteroseismology of sdB pulsators
(Table 1) or by modeling sdB+dM binaries (Table 2) are clustered near
the canonical value of $0.48~M_{\sun}$, i.e. near the mass of the
degenerate He core at helium ignition in low mass red giants.  However,
there are at least one or two other hot subdwarfs for which masses lower than
$0.4~M_{\sun}$ have also been found.

The first anomalously low mass for a hot subdwarf was found for the
eclipsing sdO+dM binary, AA~Dor, although this result continues to be
the subject of debate (\citealp{Rucinski09}; \citealp{Fleig08};
\citealp{Vuckovic08}, and references therein).  The most recent values
for the sdO mass, 0.25~$M_{\sun}$ \citep{Rucinski09} and
0.24~$M_{\sun}$ (from Fleig et al.'s values for the surface gravity,
5.30, and radius, 0.181$~R_{\sun}$) are too low for a core helium
burning star, implying that AA~Dor is on a post-RGB cooling
track, as originally suggested by \citet{Pacz80}.  This is consistent
with the fact that AA~Dor (42000~K) is much hotter than sdB stars.

\citet{Heber04,Heber05} used the MORO code to model light curves of
HS\,2333+3927, the non-eclipsing sdOB+dM binary with the largest known
reflection effect, and found two good solutions with quite different
secondary albedos, $A_{2} = 0.39$ and $A_{2} = 1.00$.  Interestingly,
their spectroscopic $\log g$ and mass--radius relations convincingly
argued that the lower albedo solution should be preferred -- the
opposite of what has been required for all other sdB+dM light curve
modeling -- resulting in a primary mass of $0.38 \pm 0.09~M_{\sun}$
for HS\,2333+3927.  However, Heber et al.\ pointed out that a mass of
$0.47~M_{\sun}$ corresponds to $\log g = 5.86$, only 0.16 dex larger
than their observed spectroscopic $\log g = 5.70$, leaving room for
doubt about the mass.  While it is clear that a non-eclipsing system
is inherently more uncertain than an eclipsing one, there are two
further pieces of evidence in favor of a lower mass for HS\,2333+3927.
Heber et al.'s gravity was derived using zero metallicity NLTE
atmospheres, and if the metallicity corrections at 36000\,K go in the
same direction as they do at several thousand degrees cooler, then any
such corrections should reduce the gravity, and therefore lower the
derived mass.  We can also corroborate their observed surface gravity
from our own independent measurements of multiple high S/N spectra
taken within 15 minutes of the minimum of the reflection effect
\citep{Green08}, similarly analyzed with zero-metal NLTE synthetic
atmospheres.  While optical spectra are not as free from the secondary
contamination as ultraviolet spectra, our derived $\log g$ of 5.70 is
nevertheless identical to Heber et al.'s value, supporting their lower
value for the mass.  (Heber et al.\ alternately suggested that
HS\,2333+3927 might be on a post-RGB cooling track, although that
would require an even lower mass of 0.29$~M_{\sun}$.)

\citet{Ostensen08} reported a very low mass ($< 0.3~M_{\sun}$) for the
eclipsing sdB, HS\,2231+2441, but their result is rather uncertain, as
it depends strongly on the spectroscopic $\log g = 5.39$, which was
determined using solar abundances.  Our independent estimate of the
gravity for this star, using the same homogeneous zero-metal NLTE
atmospheric models that we used for 2M\,1533+3927 and HS\,2333+3927,
is 5.51, consistent with a mass of $0.47~M_{\sun}$.  The true value is
presumably somewhere in between.  Further investigation is required to
better assess the sdB mass in HS\,2231+2441.

\citet{Randall07} utilized the completely different technique of
asteroseismology to derive a mass of $0.39 \pm 0.01~M_{\sun}$ for the
p-mode sdB pulsator, PG\,0911+456.  The high precision is due to the
fact that the envelope pulsations are extremely sensitive to the
surface gravity.  It turns out that any systematic metallicity
corrections would also tend to reduce the mass in this case, as well.
This is because the asteroseismic models were calculated for a fixed
temperature, the observed spectroscopic value of 31940~K, which was
once again determined by fits to synthetic zero metal NLTE
atmospheres.  There is a known degeneracy in mass {\it vs} temperature
(and gravity) for similar sdB asteroseismic solutions \citep{CF05}.
For PG\,0911+456, every 400~K decrease in the assumed effective
temperature due to metallicity corrections would lower the derived sdB
mass by about $0.01~M_{\sun}$.

Given the robustness of our light curve solution, the mass of $0.376
\pm 0.055~M_{\sun}$ for 2M\,1533+3927 appears rather firm.  Thus,
there is now significant evidence from two completely independent
observational and analytical techniques, asteroseismology and light
curve modeling in binary stars, for the existence of sdB stars with
masses around $0.38~M_{\sun}$.

Even one or two sdB stars with masses less than $0.40-0.43~M_{\sun}$,
out of about 16 whose masses are fairly well determined,
constitute an important fraction.  One such star might conceivably lie
on a post-RGB cooling track but the odds are very much against it.
For example, 2M\,1533+3927, PG\,0911+456, and HS\,2333+3927 all fall
near the extremely fast loop at the beginning of Althaus et al.'s
(2001) $0.406~M_{\sun}$ cooling track (between C and D in their
Figure~1), but the few years spent in that early phase are
insignificant compared to typical core helium burning lifetimes ($\sim
10^{8}$~yr).  The only post-RGB stars with any reasonable likelihood
of being seen at the temperatures and gravities of typical sdB stars
have masses less than $0.30~M_{\sun}$ (\citealp{ASB01}; see also
Figure~10 of \citealp{Heber04}).  The evidence therefore suggests that
sdB stars with masses near $0.38~M_{\sun}$ are bonafide core helium
burning horizontal branch stars.

The mass of PG\,0911+456 is more precisely known and therefore the
evolutionary history is more interesting. It does not now appear to be
in a binary system \citep{Randall07}, and it isn't clear why some, but
not all, single $\sim 2~M_{\sun}$ progenitors would lose their entire
envelopes.  The merger of two helium white dwarfs is not a completely
satisfactory alternative -- Han et al.'s sdB models give a lower limit
of $0.4~M_{\sun}$ for the product of such a merger -- unless some of
the mass in the two white dwarfs can somehow manage to escape during
the merger. Politano et al's (2008) common envelope merger model
predicts a lower mass limit ($\leq 0.32~M_{\sun}$), in better
agreement with the observed mass of PG\,0911+456.  Their model also
hypothesizes that since fast rotators lose more envelope mass, a
significant fraction of the envelope angular momentum would be carried
away, slowing down the star's rotation.  However, PG\,0911+456 has an
unusually low rotational velocity, less than 0.1~km~s$^{-1}$, and it
is not clear if a common envelope merger could explain the loss of
essentially all the envelope mass as well as nearly all the angular
momentum.

2M\,1533+3759 has clearly been through an initial
common envelope.  Theoretical investigations, from the first in-depth
study by \citet{SGR89} to recent work aimed specifically at binary
systems expected to produce hot subdwarfs \citep{Han02,Han03,Hu07},
indicate that helium burning cores somewhat less than $0.40~M_{\sun}$
are produced by stars with initial masses greater than about
$2~M_{\sun}$, which undergo non-degenerate helium ignition.  Of
course, 2M\,1533+3759 might still have had a degenerate helium flash
if the mass of the sdB is towards the upper end of the possible range.
Still, either way, a helium core mass less than about $0.43~M_{\sun}$
ought to have evolved from a main sequence progenitor with an initial
mass of at least $1.8-2.0~M_{\sun}$, which corresponds to a main
sequence A star \citep{BM98}.  2M\,1533+3759 therefore presents the best
observational evidence so far that stars with initial main sequence
masses this large can be sdB progenitors.  (The situation in sdB
binaries with compact companions is less clear, since mass may have
been transferred to the sdB progenitor during the evolution of the
original primary.)

Previously, the upper limit to the mass of an sdB progenitor could
only be estimated from the fact that sdB stars have not been found in
any galactic clusters younger than NGC\,188, which has an age of 6--7
Gyr and a turnoff mass of $1.1~M_{\sun}$ \citep{Meibom09}.  Small
number statistics clearly play an important role here, since there are
only two hot subdwarfs in NGC\,188, and half a dozen or so in NGC\,6791
\citep{Landsman98}, the only other old open cluster known to contain
such stars, and the majority of younger open clusters are even less
massive than these two.

Indeed, at a mass of $0.38~M_{\sun}$, 2M\,1533+3759 (and perhaps also
HS\,2333+3927, if the latter's mass is in fact less than
$0.4~M_{\sun}$) would fall at the low mass end of Han et al.'s (2003)
preferred distribution for the first common envelope ejection channel
(see their Figure~12).  The existence of a binary like 2M\,1533+3759
therefore may also provide support for Han et al's (2002, 2003)
assumption that a fraction of the ionization energy contained in the
progenitor red giant's envelope combines with the liberated
gravitational potential energy to enable the ejection of the common
envelope.  Without this extra energy, it would be more difficult to
eject the envelope around such a massive red giant and a
$0.1~M_{\sun}$ M dwarf secondary, and the two might well merge
\citep{Sandquist00}.

\section{SUBSEQUENT EVOLUTION}

We consider the possible CV scenario for the subsequent evolution of 
2M\,1533+3759. If we assume gravitational radiation is 
the only acting mechanism for angular momentum loss and the secondary 
has not evolved on this time scale, the orbital period will 
decrease until the Roche lobe comes into contact
with the secondary, initiating mass transfer and the beginning of
the cataclysmic variable (CV) stage.  The orbital period at
contact, $P_{c}$ can be calculated using Kepler's law and the fact the
ratio of the Roche lobe radius to the orbital separation is constant prior to
contact: $P_{c} = P (a_{c} / a)^{1.5} = P (R_{2} / R_{L2})^{1.5}$,
where $a_{c}$ is the orbital separation at the beginning of contact, $a$
is the current orbital separation, $R_{2} = 0.152~R_{\sun}$ is the
radius of the secondary (which is assumed not to change
significantly), and $R_{L2} = 0.276~R_{\sun}$ is the current Roche
lobe of the secondary \citet{Eggleton83}.

The resulting $P_{c}$, 0.066~d (1.6~h), will be above the mininum
orbital period (1.27 hr) for a cataclysmic variable and below the
period gap \citep{Knigge06}. If any additional mechanisms, such as
magnetic braking, have a significant effect (see \citealp{Sills00}),
the time scale for Roche lobe contact would be reduced.

\section{CONCLUSION}

The sdB star 2M\,1533+3759 is the seventh eclipsing sdB+dM binary
discovered to date.  Its orbital period of 0.16177042 $\pm$
0.00000001~d is 29\% longer than the 0.12505 day period of the next
longest eclipsing sdB+dM, BUL$-$SC16~335.  The amplitude of the reflection
effect in 2M\,1533+3759 is surprisingly strong, only about 0.05 mag
weaker than the amplitude observed in HW~Vir, in spite of the longer
orbital period and the fact that the temperatures of the primary stars
are similar.

2M\,1533+3759 is the only new sdB binary among the eclipsing systems
that were proposed to be sdB+dM by \citet{KS07} on the basis of their
narrow eclipse widths.  This result is consistent with the 2MASS
colors of other known reflection-effect sdB+dM systems, all of which
have $J-H < 0$. 2M\,1533+3759 and the archetypal HW~Vir \citep{MM86}
are the only two binaries in Kelley \& Shaw's (2007) Table~3 that have
similarly blue IR colors, and the only two that contain sdB stars.

Spectroscopic parameters 2M\,1533+3759 were derived by fitting Balmer
and helium line profiles in high S/N spectra to a grid of
zero-metallicity NLTE model atmospheres. The effective temperatures
derived from low (9\AA) and medium (1.9\AA) resolution spectra exhibit
clear variations with orbital phase.  Phase variations are much less
significant for the surface gravities, and completely negligible for
the helium abundance fraction.  Our adopted parameters for the sdB
star, $T_{\rm eff} = 29230 \pm 125$\,K, $\log g = 5.58 \pm 0.03$,
log $N$(He)/$N$(H) =$ -2.37 \pm 0.05$, were determined from medium resolution
spectra taken when the reflection effect was near minimum.  The
inferred rotational velocity has a negligible affect on the derivation
of these parameters.

Light curve modeling with the MORO code produced only one well-fitting
solution consistent with a core helium burning primary.  The system
mass ratio, $q$ ($M_{\rm 2}$/$M_{\rm 1}$), is $0.301 \pm 0.014$ and
the inclination angle, $i$, is $86.6\degr \pm 0.2\degr$.  The
robustness and precision of these numbers are due to the high
precision of the light curves and the fact that the system is
eclipsing.  Radial velocities for the sdB component were used to
derive the velocity amplitude, K$_{\rm 1}$ = $71.1 \pm
1.0$~km~s$^{-1}$, leading to component masses of $M_{\rm 1}$ = $0.376
\pm 0.055~M_{\sun}$ and $M_{\rm 2}$ = $0.113 \pm 0.017~M_{\sun}$.  The
errors in the masses are dominated by the uncertainty in $q$.  Since
the mass ratio and inclination are even more uncertain in
non-eclipsing systems, our inability to more tightly constrain the
primary mass provides a strong illustration for why useful sdB masses
from light curve modeling can usually be obtained only from eclipsing
binaries.

The orbital separation derived from the masses and the period is $a =
0.98 \pm 0.04~R_{\sun}$.  The individual radii, $R_{\rm 1}$ = $0.166
\pm 0.007~R_{\sun}$, $R_{\rm 2}$ = $0.152 \pm 0.005~R_{\sun}$ were
then calculated from the relative radii, $R_{\rm 1}/a$ and $R_{\rm
  2}/a$, determined by the light curve solution.  Both radii are
consistent with theoretical expectations, and the resulting sdB
surface gravity, $\log g = 5.57 \pm 0.07$, is completely consistent
with the adopted spectroscopic value above.

We constructed a synthetic line-blanketed spectrum to investigate
potential systematic effects caused by our use of zero metallicity
NLTE atmospheres to derive the spectroscopic parameters.  If
2M\,1533+3759 had solar abundances of C, N, O, S, and Fe in its
atmosphere, our assumption of zero metals would have overestimated the
effective temperature by about 2000\,K, and the surface gravity by
almost 0.2 dex.  Thus, the true $T_{\rm eff}$ and $\log g$ abundances
would have been about 27300\,K and 5.40, respectively.  The modeled
light curve solution at this lower temperature is only negligibly
different from our original solution, and thus the resulting system
parameters remain essentially unchanged.  However, in this case, the
calculated sdB surface gravity, $\log g = 5.57$, would be much less
consistent with the expected value of 5.40.  This suggests that the
full correction to solar metallicites assumed above is not appropriate
for 2M\,1533+3759.

An important conclusion is that there is now significant observational
evidence, from two completely independent techniques, asteroseismology
(PG\,0911+456) and modeling of eclipsing/reflection effect light
curves (2M\,1533+3759, and perhaps HS\,2333+3927), for the existence
of sdB stars with masses significantly lower than the canonical $0.48
\pm 0.02~M_{\sun}$.

2M\,1533+3759 must have formed via the first common-envelope channel,
since the companion is an M dwarf.  With a probable sdB mass in the
range $0.32-0.43~M_{\sun}$, this star is expected to have evolved from
a main sequence A star with an initial mass $> 1.8-2.0~M_{\sun}$.  The
existence of such a binary might support recent theoretical predictions
that sdB stars can be produced by such massive progenitors, including
the assumption that the ionization energy of the red giant envelope
contributes to the ejection of the common envelope
\citep{Han02,Han03}.  If the primary mass of 2M\,1533+3759 could be
measured more precisely, or if the separation between the two
components could be measured independently, this system ought to
provide a very useful observational constraint for the upper limit to
the main sequence mass of an sdB progenitor.

If 2M\,1533+3759 becomes a cataclysmic variable (CV) after orbital shrinkage due 
to gravitational radiation brings the Roche lobe into contact with the M dwarf 
secondary, its orbital period of the CV at the onset of mass transfer will be 
1.6 hours, below the CV period gap.

\acknowledgements

We acknowledge the invaluable assistance of the Steward mountain staff
at the Catalina and Kitt Peak observatories.  We are also in debt to
Bill Peters for the excellent error treatment in his linearized least
squares program, and to Roy {\O}stensen for helping to resolve the
question of NSVS\,04818255. The authors thank the referee for thoughtful 
comments that helped to improve the original manuscript. 

\clearpage
\begin{deluxetable}{lccccl}
\tablewidth{0pt}
\rotate
\tablecaption{sdB stars with masses determined by asteroseismology}
\tablehead{
\colhead{Name} &
\colhead{log $g$} &
\colhead{$T_{\rm eff}$} &
\colhead{$M_1$} &
\colhead{log $M_{\rm env}/M_*$} &
\colhead{References}\\
\colhead{} &
\colhead{(cm s$^{-2}$)} &
\colhead{(K)} &
\colhead{($M_{\sun}$)} &
\colhead{} &
\colhead{} 
}
\startdata
PG\,1047$+$003   &   5.800$\pm$0.006 &  33150$\pm$200 & 0.490$\pm$0.014 &
$-3.72$$\pm$0.11 &  \citet{Charpinet03} \\
PG\,0014$+$067   &   5.775$\pm$0.009 &  34130$\pm$370 & 0.477$\pm$0.024 &
$-4.32$$\pm$0.23 &  \citet{Charpinet05} \\
PG\,1219$+$534   &   5.807$\pm$0.006 &  33600$\pm$370 & 0.457$\pm$0.012 &
$-4.25$$\pm$0.15 &   \citet{CF05}        \\
PG\,1325$+$101   &   5.811$\pm$0.004 &  35050$\pm$220 & 0.499$\pm$0.011 &
$-4.18$$\pm$0.10 &   \citet{Charpinet06} \\
EC\,20117$-$4014 &   5.856$\pm$0.008 &  34800$\pm$2000 & 0.540$\pm$0.040 &
$-4.17$$\pm$0.08 &   \citet{Randall06}   \\
PG\,0911$+$456   &   5.777$\pm$0.002 &  31940$\pm$220 & 0.390$\pm$0.010 &
$-4.69$$\pm$0.07 &   \citet{Randall07}   \\
Feige\,48        &   5.462$\pm$0.006 &  29580$\pm$370 & 0.519$\pm$0.009 &
$-2.52$$\pm$0.06 &   \citet{VanGrootel08a} \\
BAL\,090100001   &   5.383$\pm$0.004 &  28000$\pm$1200 & 0.432$\pm$0.015 &
$-4.89$$\pm$0.14 &   \citet{VanGrootel08b} \\
PG\,1336$-$018   &   5.739$\pm$0.002 &  32740$\pm$400 & 0.459$\pm$0.005 & 
$-4.54$$\pm$0.07 &   \citet{Charpinet08}  \\
PG\,1605$+$072   &   5.226$\pm$0.005 &  32300$\pm$400 & 0.528$\pm$0.004 &
$-5.88$$\pm$0.04 &   \citet{VanS08}    \\
EC\,09582$-$1137 &   5.788$\pm$0.004 &  34805$\pm$230 & 0.485$\pm$0.011 &
$-4.39$$\pm$0.10 &   \citet{Randall09}   \\
\enddata
\end{deluxetable}

\clearpage
\begin{table}
\scriptsize
\begin{minipage}{100mm}
\caption{Currently known sdB$+$dM binaries}
\begin{tabular}{@{}lllllll@{}}
\tableline
\multicolumn{7}{c}{Reflection Effect/Eclipsing Binaries}\\ 
\tableline
Name & Alternate Name & Period &  $M_{1}$        &  $M_{2}$ & References
& Comments\\
     &                  & (day)  &  ($M_{\sun}$)  &   ($M_{\sun}$) &   &
\\
\tableline
HS\,0705+6700 &                   & 0.0956466   &   0.48    & 0.13   &
\citet{Drechsel01} &    light curve \\
PG\,1336$-$018  &   NY\,Vir         & 0.101015999 &   0.466/0.389   &
0.122/0.110   &  Vuckovic et al. 2007 &  light curve, two solutions\\
                &                   &             &   0.459   & --
& \citet{Charpinet08} & asteroseismology \\
NSVS\,14256825  &   J\,2020+0437    &  0.1104     &   0.46    & 0.21
& \citet{Wils07} &      no spectroscopy \\
HS\,2231+2441   &                   & 0.11058798  &   $< 0.3$ & --
& \citet{Ostensen08} & uncertain $\log g$ \\
PG\,1241$-$084  &   HW\,Vir         & 0.11676195  &   0.485   & 0.142
& \citet{Lee09}   &    light curve \\
BUL--SC16 335   &                   & 0.125050278 &   --      & --
& \citet{Polubek07}  &  \\
2M\,1533+3759   &   NSVS\,07826147  & 0.16177042  &   0.377   & 0.113
& this paper         &  light curve \\
\tableline
\multicolumn{7}{c}{Reflection Effect/Non-Eclipsing Binaries}\\ 
\tableline
PG\,1017$-$086  &   XY\,Sex         & 0.073        & --       & --
& \citet{Maxted02} & \\
HS\,2333+3927  &                   & 0.1718023    & 0.38     & 0.29    &
\citet{Heber05} &  light curve \\
PG\,1329+159  &   Feige\,81, PB\,3963   & 0.249699     & --       & --
& \citet{Maxted04} &\\
                &      & 0.249702      & --       & --      &
\citet{Green04} &\
 \\
2M\,1926+3720    &   KBS~13         & 0.2923       & --       & --
& \citet{For08}  & \\
PG\,1438$-$029  &                  & 0.33579       & --       & --
& \citet{Green04} & \\
HE~0230$-$4323  &                  & 0.4515       & --       & --      &
\citet{Koen07} & \\
\tableline
\end{tabular}
\end{minipage}
\end{table}

\clearpage
\begin{deluxetable}{cclrcccc}
\tablewidth{0pt}
\rotate
\tablecaption{NSVS sources identified by Kelley \& Shaw (2007) as
  potential sdB stars} 
\tablehead{
\colhead{NSVS ID$^{\rm a}$} &
\colhead{$V^{\rm a}$} &
\colhead{Period$^{\rm a}$} &
\colhead{$J-H^{\rm b}$} &
\colhead{RA (J2000)$^{\rm b}$} &
\colhead{DEC (J2000)$^{\rm b}$} &
\colhead{Spectral Type$^{\rm c}$} &
\colhead{Comments}\\
\colhead{} &
\colhead{(mag)} &
\colhead{(day)} &
\colhead{(mag)} &
\colhead{(h m s)} &
\colhead{($\degr~\arcmin~\arcsec$)} &
\colhead{} &
\colhead{} 
}
\startdata
02335765  & 10.69 &  9.744983  & $0.224$  &  06:31:02.7 & $+$61:14:29 &
F2--F5  &    \\
03259747  & 11.22 &  1.239805  & $0.274$  &  20:57:27.7 & $+$56:46:06 &
F9--G0  &    \\
04818255  & 12.10 &  0.1600359 & $0.392$  &  08:40:58.4 & $+$39:56:28 &
G0      &   late-type eclipsing binary star        \\
        &       &            & $0.343$  &  08:41:00.2 & $+$39:55:54 &
F9--G0  &   star nearest to NSVS coords  \\
04963674  & 10.63 &  3.6390769 & $0.297$  &  11:03:36.4 & $+$41:36:02 &
F9--G0  &    \\
07826147  & 13.61 &  0.16177   & $-0.084$ &  15:33:49.4 & $+$37:59:28 &
sdB     &   2M\,1533$+$3759; FBS\,1531$+$381     \\
08086052  & 11.94 &  1.853631  & $0.255$  &  18:03:11.9 & $+$32:11:14 &
F8--F9  &    \\
09729507  & 11.77 &  4.740887  & $0.094$  &  06:05:18.4 & $+$20:44:32 &
A0--A2  &    \\
15864165  & 12.65 &  1.232349  & $0.111$  &  11:05:06.6 & $-$09:01:33 &
A6--A7  &    \\
15972828  & 11.21 &  0.116719  & $-0.119$ &  12:44:20.2 & $-$08:40:16 &
sdB     &   HW Vir    \\
\enddata
\tablenotetext{a}{Table 3 of Kelley \& Shaw (2007).}
\tablenotetext{b}{2MASS All-Sky Point Source Catalog (Skrutskie et
  al. 2006).}
\tablenotetext{c}{Steward 2.3~m spectra.}
\end{deluxetable}

\clearpage
\begin{deluxetable}{lcccc}
\tablewidth{0pt}
\tablecaption{Low resolution 2.3~m spectra}
\tablehead{
\colhead{UT Date} &
\colhead{HJD at midpoint} &
\colhead{Exp Time} &
\colhead{S/N} &
\colhead{Orbital} \\
\colhead{} &
\colhead{(2450000+)} &
\colhead{(s)} &
\colhead{} &
\colhead{Phase} 
}
\startdata
27 Jun 2005 &   3548.82037 &      550 &    165 &   0.72 \\
30 Dec 2007 &   4465.04391 &      480 &    174 &   0.44 \\
31 Dec 2007 &   4466.03402 &      400 &    161 &   0.56 \\
19 Jan 2008 &   4485.02730 &      490 &    162 &   0.97 \\
19 Sep 2008 &   4728.61983 &      450 &    179 &   0.76 \\
\enddata
\end{deluxetable}

\clearpage
\begin{deluxetable}{lcccrcc}
\tablewidth{0pt}
\tablecaption{Medium resolution 2.3~m spectra and the derived radial
  velocities} 
\tablehead{
\colhead{UT Date} &
\colhead{HJD at midpoint} &
\colhead{Exp Time} &
\colhead{S/N} &
\colhead{$V$} &
\colhead{$V_{\rm err}$} &
\colhead{Orbital} \\
\colhead{} &
\colhead{(2450000+)} &
\colhead{(s)} &
\colhead{} &
\colhead{(km s$^{-1}$)} &
\colhead{(km s$^{-1}$)} &
\colhead{Phase} 
}
\startdata
19 Feb 2008 &   4516.02529 &    750 &  83.5  &    $   27.15$ &  4.99  &
0.58 \\
18 Mar 2008 &   4543.99112 &    550 &  80.7  &    $  -24.90$ &  5.54  &
0.46 \\
18 Mar 2008 &   4544.01329 &    550 &  80.0  &    $   30.54$ &  4.73  &
0.59 \\
27 Mar 2008 &   4552.97753 &    500 &  47.3  &    $   -9.29$ &  6.68  &
0.01 \\
17 Apr 2008 &   4573.93042 &    600 &  68.9  &    $   15.07$ &  4.80  &
0.53 \\
18 Apr 2008 &   4574.94859 &    550 &  61.3  &    $   68.75$ &  5.65  &
0.82 \\
25 Apr 2008 &   4581.88679 &    500 &  77.5  &    $   73.37$ &  5.03  &
0.71 \\
25 Apr 2008 &   4581.98355 &    625 &  79.0  &    $  -69.29$ &  3.69  &
0.31 \\
26 Apr 2008 &   4582.87608 &    550 &  77.9  &    $   55.20$ &  3.76  &
0.83 \\
26 Apr 2008 &   4582.96181 &    500 &  81.9  &    $  -57.65$ &  3.96  &
0.36 \\
05 Feb 2009 &   4868.02541 &    525 &  69.7  &    $   -2.92$ &  4.54  &
0.51 \\
14 Mar 2009 &   4904.83567 &    725 &  89.0  &    $  -34.43$ &  4.18  &
0.05 \\
14 Mar 2009 &   4904.84734 &    575 &  78.8  &    $  -50.58$ &  3.84  &
0.13 \\
14 Mar 2009 &   4904.85772 &    550 &  75.2  &    $  -68.20$ &  4.23  &
0.19 \\
14 Mar 2009 &   4904.86738 &    550 &  77.9  &    $  -78.37$ &  4.74  &
0.25 \\
14 Mar 2009 &   4904.87654 &    550 &  80.0  &    $  -71.28$ &  4.86  &
0.31 \\
15 Mar 2009 &   4905.83299 &    600 &  71.6  &    $  -66.07$ &  4.46  &
0.22 \\
15 Mar 2009 &   4905.84391 &    600 &  78.2  &    $  -76.04$ &  5.03  &
0.29 \\
15 Mar 2009 &   4905.89487 &    550 &  79.0  &    $   36.78$ &  5.24  &
0.60 \\
15 Mar 2009 &   4905.90420 &    500 &  75.8  &    $   60.08$ &  4.26  &
0.66 \\
15 Mar 2009 &   4905.91322 &    500 &  74.7  &    $   64.20$ &  4.29  &
0.71 \\
15 Mar 2009 &   4905.92344 &    500 &  74.0  &    $   66.27$ &  4.19  &
0.78 \\
15 Mar 2009 &   4905.93239 &    500 &  73.5  &    $   63.25$ &  4.07  &
0.83 \\
15 Mar 2009 &   4905.94190 &    500 &  71.5  &    $   43.94$ &  3.87  &
0.89 \\
15 Mar 2009 &   4905.95137 &    575 &  73.7  &    $   16.80$ &  3.93  &
0.95 \\
15 Mar 2009 &   4905.96212 &    700 &  59.1  &    $  -19.75$ &  4.70  &
0.02 \\
15 Mar 2009 &   4905.97491 &    625 &  78.1  &    $  -38.16$ &  4.60  &
0.10 \\
16 Mar 2009 &   4906.82916 &    575 &  89.5  &    $  -52.76$ &  3.39  &
0.38 \\
16 Mar 2009 &   4906.86126 &    525 &  86.6  &    $   27.75$ &  4.86  &
0.57 \\
16 Mar 2009 &   4906.87078 &    490 &  79.2  &    $   44.75$ &  3.83  &
0.63 \\
16 Mar 2009 &   4906.88020 &    490 &  80.5  &    $   62.30$ &  4.13  &
0.69 \\
16 Mar 2009 &   4906.88876 &    490 &  81.6  &    $   63.09$ &  4.88  &
0.74 \\
16 Mar 2009 &   4906.90777 &    490 &  82.3  &    $   51.32$ &  3.20  &
0.86 \\
16 Mar 2009 &   4906.91653 &    490 &  83.3  &    $   35.74$ &  4.19  &
0.92 \\
16 Mar 2009 &   4906.92530 &    490 &  70.8  &    $   19.09$ &  4.43  &
0.97 \\
16 Mar 2009 &   4906.93541 &    650 &  82.2  &    $  -22.00$ &  4.08  &
0.03 \\
16 Mar 2009 &   4906.94885 &    650 &  92.3  &    $  -48.09$ &  3.14  &
0.12 \\
16 Mar 2009 &   4906.97131 &    575 &  87.4  &    $  -74.59$ &  4.20  &
0.25 \\
\enddata
\end{deluxetable}

\clearpage
\begin{deluxetable}{lcccc}
\tablewidth{0pt}
\tablecaption{Photometric observations at the Steward Observatory 1.55~m 
  Mt.\,Bigelow telescope}
\tablehead{
\colhead{UT Date} &
\colhead{Start HJD} &
\colhead{End HJD} &
\colhead{Filter} &
\colhead{Exp time} \\
\colhead{} &
\colhead{(2450000+)} &
\colhead{(2450000+)} &
\colhead{} &
\colhead{(s)} 
}
\startdata
 27 Feb 2008 & 4523.879786 & 4523.982705 & B,R & 30,25 \\
 28 Feb 2008 & 4524.943268 & 4525.031564 & B,R & 30,25 \\
 06 Mar 2008 & 4531.902243 & 4532.025496 & B,R & 30,25 \\
 07 Mar 2008 & 4532.896078 & 4533.016714 & B,R & 30,25 \\
 10 Mar 2008 & 4535.898112 & 4536.025827 & B,R & 30,25 \\
 11 Mar 2008 & 4536.942407 & 4537.022719 & B,R & 30,25 \\
 29 Mar 2008 & 4554.843844 & 4555.016093 & B,R & 30,25 \\
 12 Apr 2008 & 4568.787329 & 4568.974478 & V,I & 30,45 \\
 13 Apr 2008 & 4569.831345 & 4569.994764 & V,I & 30,45 \\
 26 Apr 2008 & 4582.818149 & 4582.981342 & V,I & 30,45 \\
 27 Apr 2008 & 4583.752365 & 4583.926433 & B,R & 35,30 \\
 22 Jun 2008 & 4639.674751 & 4639.710198 & B,R & 35,30 \\
 28 Mar 2009 & 4639.674751 & 4639.710198 & B,R & 30,25 \\
\enddata
\end{deluxetable}

\clearpage
\begin{deluxetable}{lcrccr}
\tablewidth{0pt}
\tablecaption{Times of minima of 2M\,1533$+$3759}
\tablehead{
\colhead{Mid Eclipse} &
\colhead{Error} &
\colhead{Epoch} &
\colhead{Type} &
\colhead{Filter} &
\colhead{$O-C$} \\
\colhead{(HJD 2450000+)} &
\colhead{} &
\colhead{} &
\colhead{} &
\colhead{} &
\colhead{(s)} 
}
\startdata
 4523.93875 & $2.5\times10^{-5}$ & $-0.5 $  & sec. & $R$  & $ 7.2$   \\
 4524.99017 & $1.5\times10^{-5}$ & $6.0  $  & pri. & $R$  & $-0.4$   \\
 4531.94631 & $1.5\times10^{-5}$ & $49.0 $  & pri. & $R$  & $ 0.6$   \\
 4532.91693 & $1.5\times10^{-5}$ & $55.0 $  & pri. & $R$  & $ 0.4$   \\
 4532.99788 & $2.5\times10^{-5}$ & $55.5 $  & sec. & $R$  & $ 6.0$   \\
 4535.90970 & $2.5\times10^{-5}$ & $73.5 $  & sec. & $R$  & $ 1.9$   \\
 4535.99054 & $1.5\times10^{-5}$ & $74.0 $  & pri. & $R$  & $-2.0$   \\
 4536.96115 & $1.5\times10^{-5}$ & $80.0 $  & pri. & $R$  & $-3.1$   \\
 4554.91769 & $1.5\times10^{-5}$ & $191.0$  & pri. & $R$  & $-1.1$   \\
 4554.99860 & $2.5\times10^{-5}$ & $191.5$  & sec. & $R$  & $ 1.0$   \\
 4568.82995 & $1.5\times10^{-5}$ & $277.0$  & pri. & $V$  & $-0.8$   \\
 4568.91082 & $2.5\times10^{-5}$ & $277.5$  & sec. & $V$  & $-2.1$   \\
 4569.88151 & $2.5\times10^{-5}$ & $283.5$  & sec. & $V$  & $ 3.7$   \\
 4569.96228 & $1.5\times10^{-5}$ & $284.0$  & pri. & $V$  & $-6.2$   \\
 4582.82312 & $2.5\times10^{-5}$ & $363.5$  & sec. & $V$  & $ 1.7$   \\
 4582.90399 & $1.5\times10^{-5}$ & $364.0$  & pri. & $V$  & $ 0.3$   \\
 4583.79377 & $2.5\times10^{-5}$ & $369.5$  & sec. & $R$  & $ 4.0$   \\
 4583.87460 & $1.5\times10^{-5}$ & $370.0$  & pri. & $R$  & $-0.7$   \\
 4639.68546 & $1.5\times10^{-5}$ & $715.0$  & pri. & $R$  & $ 4.8$   \\
 4918.90113 & $1.5\times10^{-5}$ & $2441.0$ & pri. & $R$  & $-2.1$   \\
 4918.98208 & $2.5\times10^{-5}$ & $2441.5$ & sec. & $R$  & $ 3.5$   \\
\enddata
\end{deluxetable}

\clearpage
\begin{table}
\scriptsize
\centering
\begin{minipage}{100mm}
\caption{Light curve solution for 2M\,1533+3759 and goodness of fit.}
\begin{tabular}{@{}ll@{}}
\tableline
  Fixed Parameters     &   Value \\
\tableline
 $\beta_{1}^{a}$    & 1.0 \\
 $\beta_{2}^{a}$    & 0.32 \\
 $A_{1}^{b}$       & 1.0  \\
 $x_{1}(B)^{c}$    & 0.305 \\
 $x_{1}(V)^{c}$     & 0.274 \\
 $x_{1}(R)^{c}$     & 0.229 \\
 $x_{1}(I)^{c}$    & 0.195 \\
 $\delta_{2}^{d}$   & 0.0 \\
 $l_{3}(B,V,R,I)^{e}$ &   0.0 \\
\tableline
  Adjusted Parameters & Value \\
  \tableline
  $i$              &   $86.6\degr \pm 0.2\degr$      \\
  $A_{2}^{b}$       &   $2.0 \pm 0.2$  \\
  $q (M_{2}/M_{1})$ &   $0.301 \pm 0.014$                   \\
  $\Omega_{1}^f$    &   $6.049 \pm 0.230$  \\
  $\Omega_{2}^f$    &   $3.305 \pm 0.098$  \\
  $\delta_{1}^d$    &   $0.035 \pm 0.043$   \\
  $T_{\rm eff} (1)$    &   $30400 \pm 500$                  \\
  $T_{\rm eff} (2)$    &   $3100 \pm 600$                   \\
  $x_{2}(B)^c$      &   $0.83 \pm 0.17$  \\
  $x_{2}(V)^c$      &   $0.91 \pm 0.09$  \\
  $x_{2}(R)^c$      &   $0.95 \pm 0.05$  \\
  $x_{2}(I)^c$      &   $1.00 \pm 0.02$  \\
  $L{1}(B)^g$       &   $0.99996 \pm 0.00004$                  \\
  $L{1}(V)^g$       &   $0.99978 \pm 0.00017$                  \\
  $L{1}(R)^g$       &   $0.99941 \pm 0.00043$                  \\
  $L{1}(I)^g$       &   $0.99821 \pm 0.00116$                  \\
\tableline
  Fractional Roche Radii$^{h}$ & Value     \\
  \tableline
  $r_{1}$(pole)  &  $0.168 \pm  0.003$\\
  $r_{1}$(point) &  $0.169 \pm  0.003$\\
  $r_{1}$(side)  &  $0.168 \pm  0.002$\\
  $r_{1}$(back)  &  $0.169 \pm  0.002$\\
  $r_{2}$(pole)  &  $0.153 \pm  0.001$\\
  $r_{2}$(point) &  $0.154 \pm  0.004$\\
  $r_{2}$(side)  &  $0.154 \pm  0.001$\\
  $r_{2}$(back)  &  $0.157 \pm  0.003$\\
\tableline
  Standard Deviation  &    \\
  \tableline
  $\sigma_{B}$ & 0.0072 \\
  $\sigma_{V}$ & 0.0061 \\
  $\sigma_{R}$ & 0.0069 \\
  $\sigma_{I}$ & 0.0080 \\
\tableline
\end{tabular}
\\$^{a}$ Gravity darkening exponent.\\
$^{b}$ Bolometric albedo.\\
$^{c}$ Limb darkening coefficient.\\
$^{d}$ Radiation pressure parameter.\\
$^{e}$ Fraction of third light at maximum.\\
$^{f}$ Roche surface potential.\\
$^{g}$ Relative luminosity, $L_{1}/(L_{1}+L_{2})$.\\
$^{h}$ In units of separation of mass centers.\\
\end{minipage}
\end{table}

\clearpage
\begin{deluxetable}{ll}
\tablewidth{0pt}
\tablecaption{Fundamental parameters of 2M\,1533$+$3975}
\tablehead{
\colhead{Parameter} &
\colhead{Value} 
}
\startdata
  $T_{\rm eff_{1}}$ (K)           & $29230 \pm 125$ \\
  $\log g$  (cm s$^{-2}$)         & $5.58 \pm 0.03$ \\
  log $N$(He)/$N$(H)                    & $-2.37 \pm 0.05$ \\
  Period (days)                   & $0.16177042 \pm  0.00000001$\\
  $T_{0}$ (days)                  & $2454524.019552 \pm 0.000009$    \\
  $K_{1}$ (km s$^{-1}$)           & $71.1 \pm 1.0$   \\
  $\gamma$ (km s$^{-1}$)          & $-3.4 \pm 5.2$  \\
  $M_{1}$ ($M_{\sun}$)            & $0.376 \pm 0.055$ \\
  $M_{2}$ ($M_{\sun}$)            & $0.113 \pm 0.017$ \\
  $a$ ($R_{\sun}$)                & $0.98 \pm 0.04$    \\
  $R_{1}$ ($R_{\sun}$)            & $0.166 \pm 0.007$ \\
  $R_{2}$ ($R_{\sun}$)            & $0.152 \pm 0.005$    \\
  $T_{\rm eff_{2}}$ (K)           & $3100 \pm 600$ \\
  $V_{\rm rot_1}$ (km s$^{-1}$) & $52 \pm 2$   \\
  $L_1$ ($L_{\sun}$)              & $18.14 \pm 1.84$ \\
  $M_{\rm V_1}$                   & $4.57 \pm 0.21$ \\
  $d$ (pc)                        & $644 \pm 66$  \\   
\enddata
\end{deluxetable}

\clearpage
\bibliographystyle{apj}
\bibliography{apj-jour,bibliographie}

\clearpage
\centerline{\bf{FIGURE CAPTIONS}}

\noindent Fig.\ 1 ---  Flux-calibrated 2M\,1533+3759 spectrum compared
to the bluest and reddest non-sdB spectra from Table 3.

\noindent Fig.\ 2 ---  Finder chart for 2M\,1533$+$3759. The solid
circle in the center of the chart is 2M\,1533$+$3759. The dashed circles
are the adopted reference stars.

\noindent Fig.\ 3 ---  Radial velocity solution for 2M\,1533+3759 as a
function of orbital phase, superimposed on the observed velocities. The
velocity amplitude and systemic velocity are $K_{\rm 1} = 71.1 \pm
1.0$~km~s$^{-1}$ and $\gamma = -3.4 \pm 5.2$~km~s$^{-1}$.

\noindent Fig.\ 4 ---  Derived gravities (above) and effective
temperatures (below) as a function of orbital phase, from fits to
Balmer and helium lines in 2M\,1533+3759.

\noindent Fig.\ 5 ---  Fits of the Balmer and helium lines in the
combined 2M\,1533+3759 minimum light spectrum to synthetic zero
metallicity NLTE line profiles. 

\noindent Fig.\ 6 ---  The observed light curves superimposed onto the
calculated theoretical light curves (solid red lines).  The $VRI$ light curves
are each offset by a constant with respect to the $B$ light curve.

\noindent Fig.\ 7 ---  Snapshots of 2M\,1533+3759 at various orbital
phases, as viewed from an inclination angle of $86.6\degr$. Left column,
top to bottom: phase 0.00 (primary eclipse), 0.04, 0.25 and 0.47.
Right column, top to bottom: phase 0.50 (secondary eclipse), 0.53, 0.75
and 0.97.

\clearpage
\begin{figure}[p]
\plotone{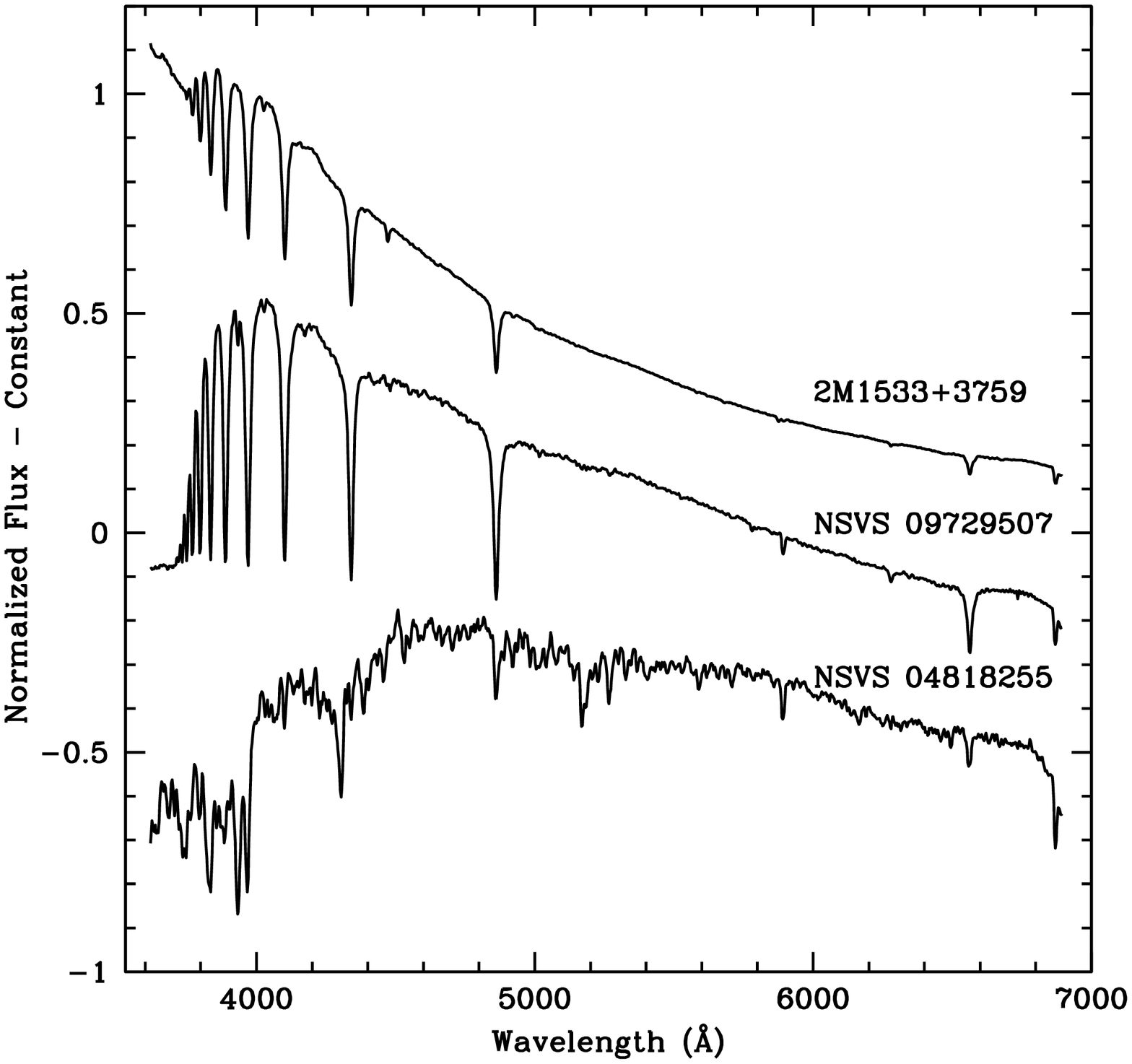}
\begin{flushright}
Figure 1
\end{flushright}
\end{figure}

\clearpage
\begin{figure}[p]
\plotone{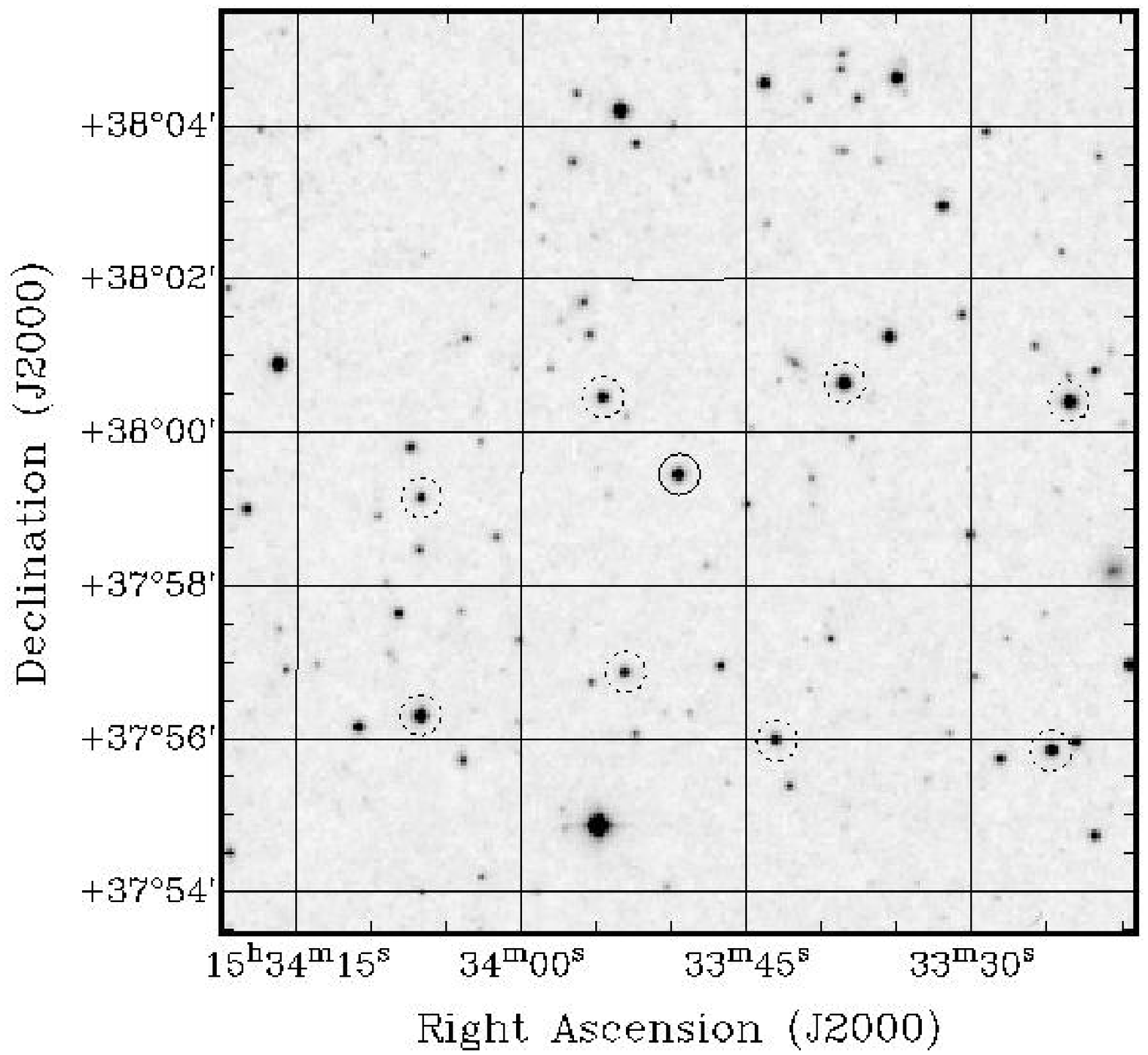}
\begin{flushright}
Figure 2
\end{flushright}
\end{figure}

\clearpage
\begin{figure}[p]
\plotone{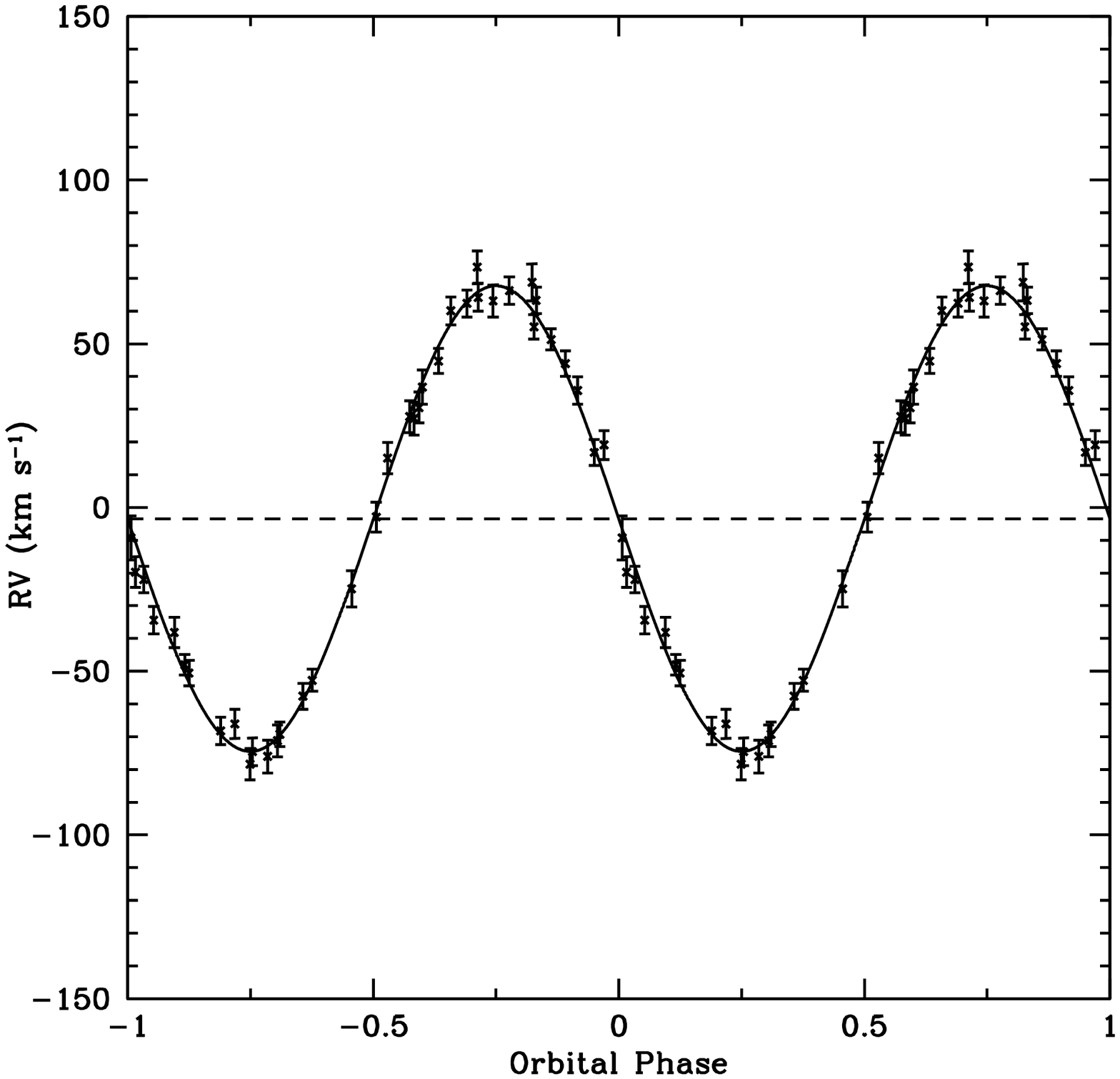}
\begin{flushright}
Figure 3
\end{flushright}
\end{figure}

\clearpage
\begin{figure}[p]
\plotone{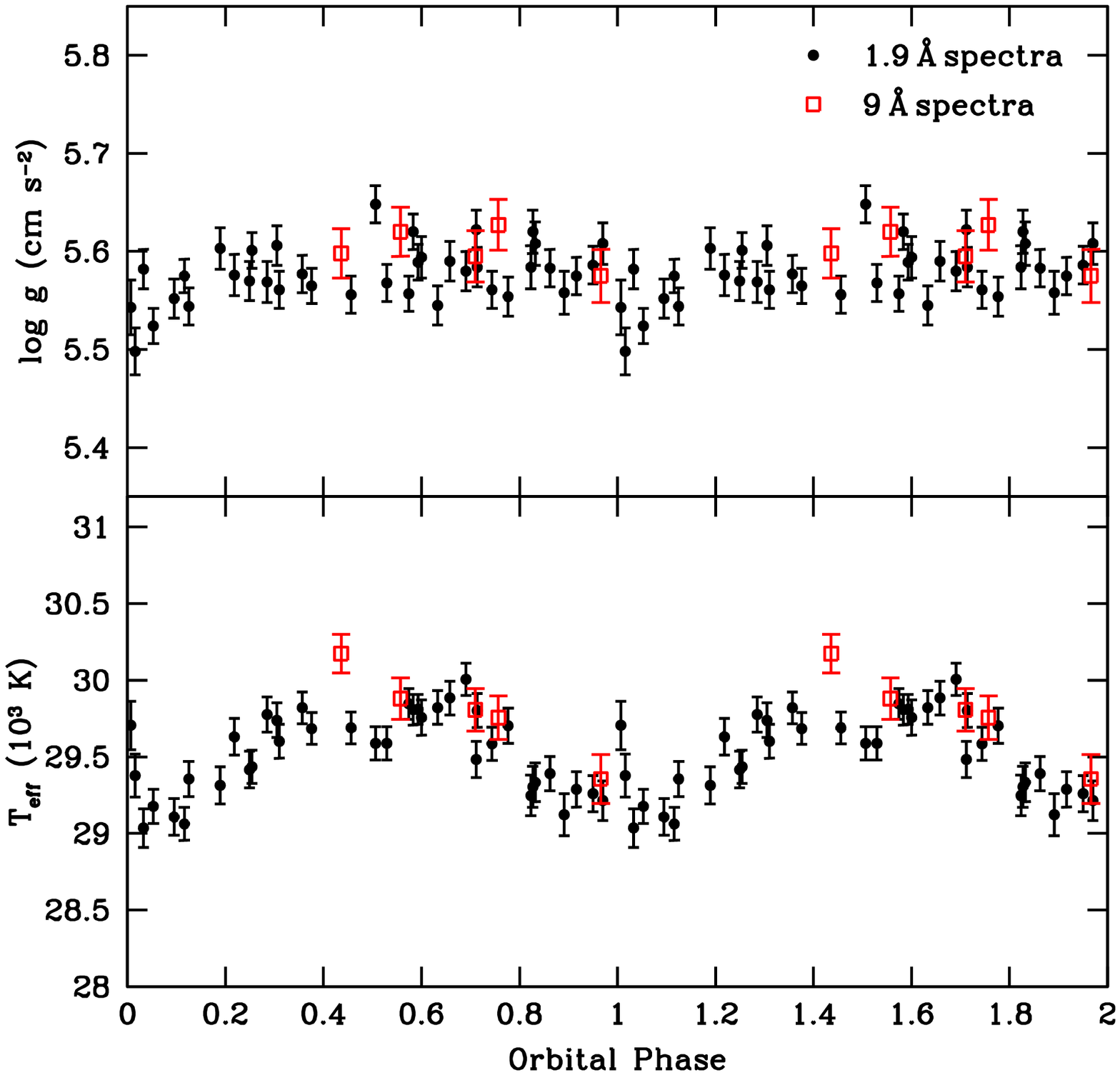}
\begin{flushright}
Figure 4
\end{flushright}
\end{figure}

\clearpage
\begin{figure}[p]
\plotone{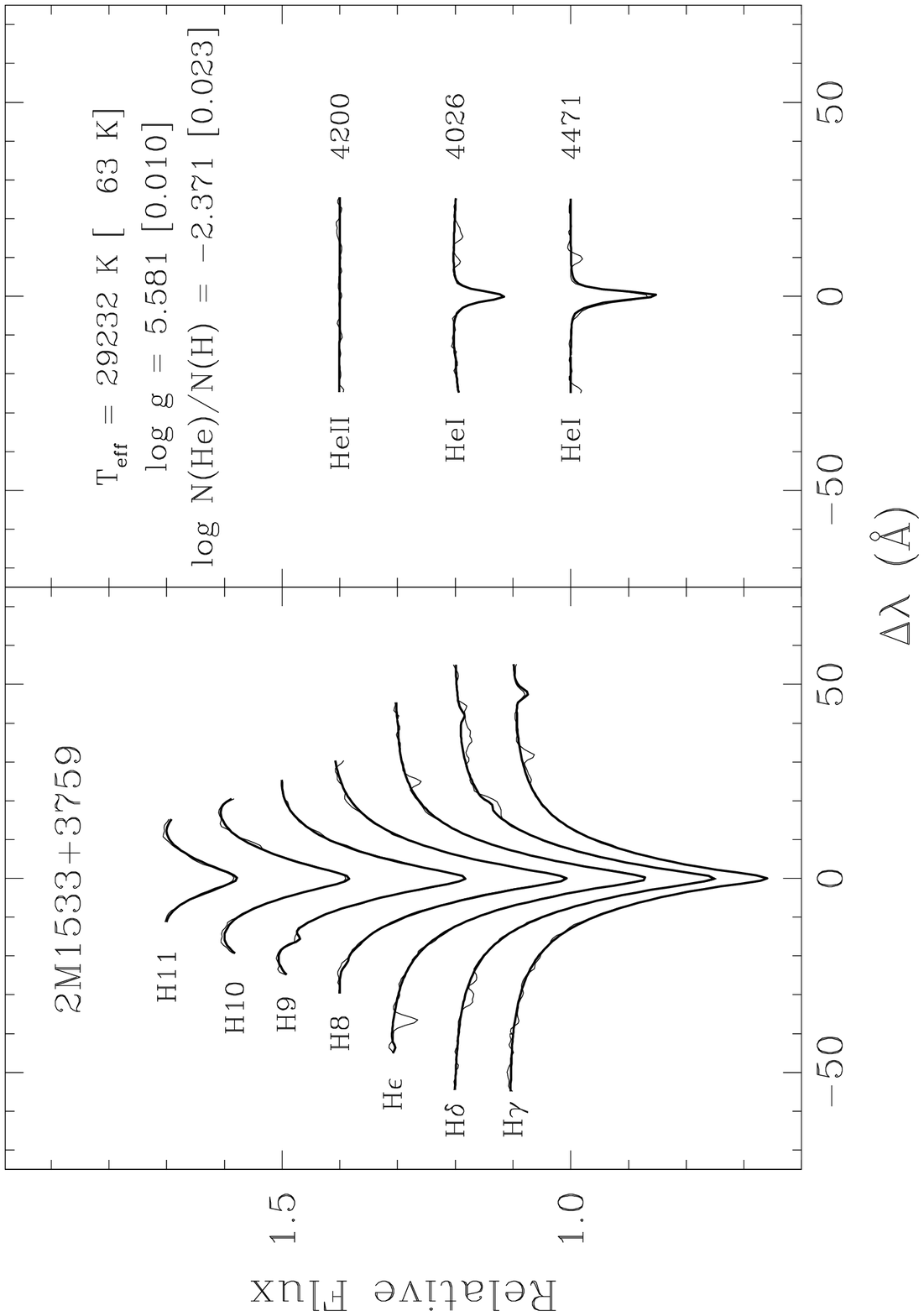}
\begin{flushright}
Figure 5
\end{flushright}
\end{figure}

\clearpage
\begin{figure}[p]
\plotone{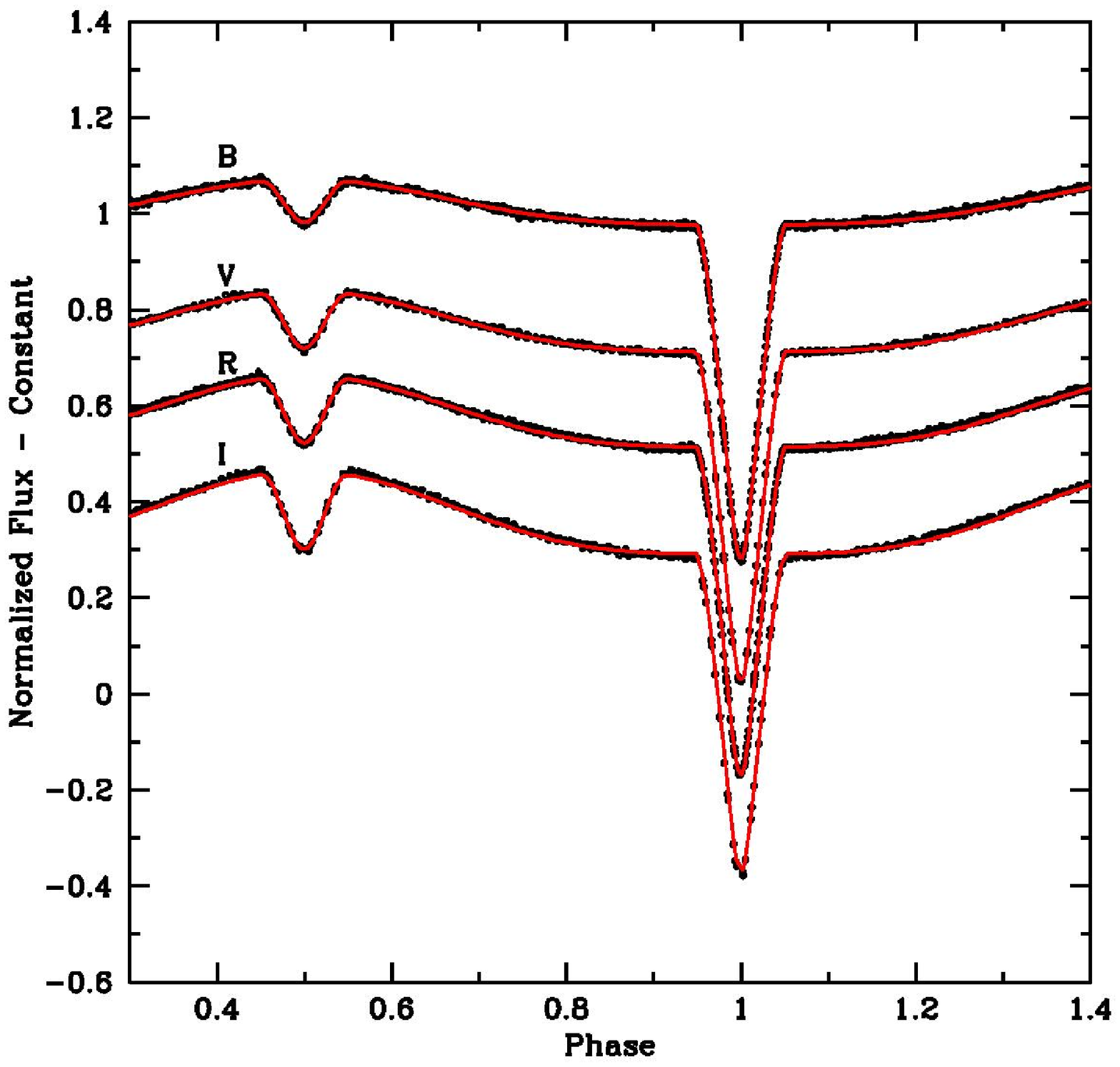}
\begin{flushright}
Figure 6
\end{flushright}
\end{figure}

\clearpage
\begin{figure}[p]
\plotone{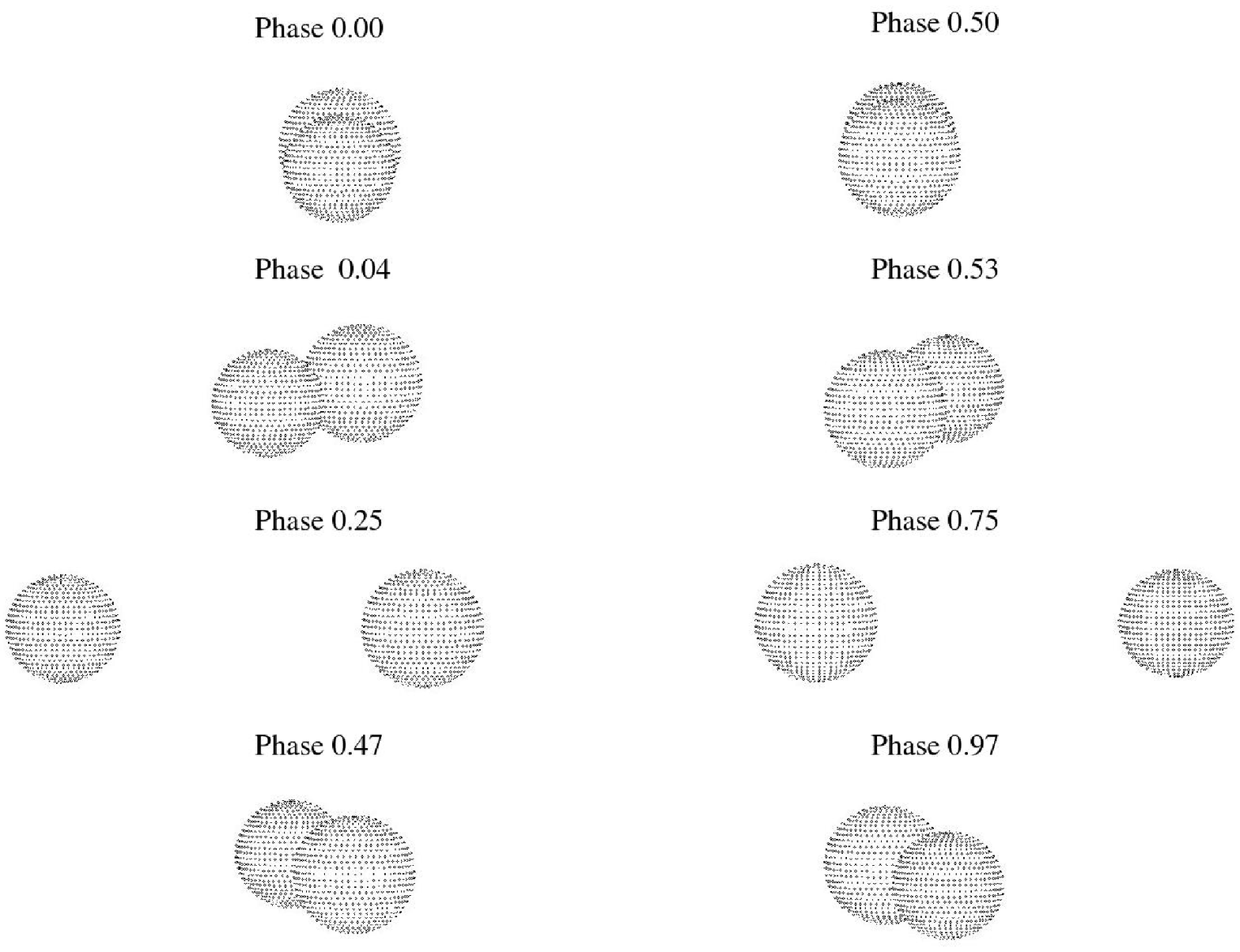}
\begin{flushright}
Figure 7
\end{flushright}
\end{figure}

\end{document}